\documentclass[fleqn,usenatbib]{mnras}

\usepackage[T1]{fontenc}

\DeclareRobustCommand{\VAN}[3]{#2}
\let\VANthebibliography\thebibliography
\def\thebibliography{\DeclareRobustCommand{\VAN}[3]{##3}\VANthebibliography}

\usepackage{amsmath}
\usepackage{amssymb}
\usepackage{bm}
\usepackage{graphicx}
\usepackage{multirow,colortbl}
\usepackage{verbatim}
\usepackage{enumitem}
\usepackage{ulem} 

\usepackage[svgnames]{xcolor}

\title[Satellite Velocity Functions]{The Milky Way satellite velocity function is a sharp probe of small-scale structure problems} 

\author[S. Y. Kim \& A. H. G. Peter ]{
Stacy Y. Kim$^{1,2,3}$\thanks{E-mail: kim.4905@osu.edu (SYK)}
and Annika H.~G. Peter$^{2,3,4}$
\\
$^{1}$Department of Physics, University of Surrey, Guildford, GU2 7XH, United Kingdom \\
$^{2}$Department of Astronomy, The Ohio State University, 140 W. 18th Ave., Columbus, OH 43210, USA \\
$^{3}$Center for Cosmology and AstroParticle Physics (CCAPP), The Ohio State University, 191 W. Woodruff Ave., Columbus, OH 43210, USA \\
$^{4}$Department of Physics, The Ohio State University, 191 W. Woodruff Ave., Columbus, Ohio 43210, USA}

\date{Accepted XXX. Received YYY; in original form ZZZ}

\pubyear{2021}

\usepackage{newtxtext,newtxmath}

\begin{document}
\label{firstpage}
\pagerange{\pageref{firstpage}--\pageref{lastpage}}
\maketitle

\begin{abstract}
Twenty years ago, the mismatch between the observed number of Milky Way satellite galaxies and the predicted number of cold dark matter (CDM) subhalos was dubbed the ``missing satellites problem".  Although mostly framed since in terms of satellite counts in luminosity space, the missing satellites problem was originally posed in velocity space.  Importantly, the stellar velocity dispersion function encodes information about the density profile of satellites as well as their abundance.  In this work, we completeness correct the MW satellite stellar velocity dispersion function down to its ultrafaint dwarfs ($L \gtrsim 340$ L$_\odot$). Our most conservative completeness correction is in good agreement with a simple CDM model in which massive, classical satellites (M$_{\rm 200} \gtrsim 5 \times 10^8~$M$_\odot$) have baryon-driven cores, while lower mass, ultrafaint satellites inhabit cuspy halos that are not strongly tidally stripped. Tidal destruction of satellites by the MW's disk must be minimal, otherwise the completeness-corrected velocity function exceeds any plausible CDM prediction---a ``too many satellites" problem.  We rule out non-core-collapsing self-interacting dark matter models with a constant cross section $\gtrsim$ 0.1 cm$^2$/g.  Constraints on warm dark matter are stronger than those based on the luminosity function due to its additional sensitivity to subhalo central densities, which suppresses number counts by up to an additional 30\%.  A thermal relic mass $\gtrsim$ 6 keV is preferred. Reducing uncertainties on stellar velocity dispersion measurements and the amount of tidal stripping experienced by the faintest dwarfs is key to determining the severity of the too many satellites problem.
\end{abstract}

\begin{keywords}
cosmology -- dark matter -- methods: numerical
\end{keywords}



\section{Introduction}
\label{sec:introduction}

While the cold dark matter (CDM) paradigm successfully explains a diverse host of observations at ``large'', cosmological scales---the scales of galaxies or larger---two fundamental predictions of CDM have met serious challenges at ``small'' scales \citep[e.g.,][]{2017ARA&A..55..343B,2018PhR...761....1B}.  Firstly, CDM predicts that the hierarchical, self-similar structure of the universe extends from galaxy cluster scales (10$^{15}$ M$_\odot$) down to Earth-mass scales \citep[10$^{-6}$ M$_\odot$;][]{2001PhRvD..64h3507H,2005JCAP...08..003G,2005PhRvD..71j3520L,2006PhRvL..97c1301P}.  However, prior to the 2000s, the observed satellite galaxies of the Milky Way (MW) numbered about an order of magnitude fewer than predicted by CDM simulations---a discrepancy that became known as the missing satellites problem.  Since then, observational surveys of satellite galaxies and non-luminous subhalos have yielded claims both in the direction that there are too few satellites \citep{1999ApJ...522...82K, 1999ApJ...524L..19M, 2015A&A...574A.113P} and too many \citep{2002ApJ...572...25D,2008ApJ...688..277T, 2014MNRAS.442.2017V,2018PhRvL.121u1302K,2019MNRAS.tmp.1496K}.

Secondly, simulations with CDM predict that dark matter halos have highly dense central regions that are sharply peaked at the center, or are ``cuspy'' \citep{1996ApJ...462..563N}.  While galaxies with sufficiently high densities have been observed \citep[e.g.][]{2020A&A...633A..36M}, many (if not most) galaxies appear to have lower central densities, including some that are consistent with a flat, or ``cored'' density profile \citep{2010ApJ...717L..87W, 2011ApJ...741L..29K, 2011MNRAS.415L..40B, 2012MNRAS.425.2817F, 2013ApJ...765...25N,2019MNRAS.484.1401R,2019ApJ...887...94R,2020ApJ...904...45H}.  Originally called the ``core-cusp'' problem, it is now sometimes called the ``density'' problem or reframed as a ``diversity'' conundrum
due to the rich range of observed galaxy rotation curves \citep[e.g.,][]{2005ApJ...621..757S,2008ApJ...676..920K,2015MNRAS.452.3650O,2016MNRAS.462.3628R,2016MNRAS.456.3542T,2019ApJ...887...94R,2019PhRvX...9c1020R}.

Both baryonic physics and novel dark-matter physics have been invoked to ``solve'' these small-scale discrepancies.  The cosmological simulations that informed the early theoretical predictions only included (cold) dark matter.  Increases in computational power and sophisticated numerical tools gave way to simulations that included increasingly sophisticated baryonic physics and/or novel dark matter physics.  Combined with advances in survey capabilities, these small-scale problems of CDM look different today than when originally formulated twenty years ago.

The discovery and interpretation of new data and new simulations have flipped the missing satellites problem on its head, leading to a potential ``too many satellites'' problem \citep{2018PhRvL.121u1302K,2019MNRAS.tmp.1496K}. Beginning with the Sloan Digital Sky Survey \citep{york2000}, wide-field surveys have revealed dozens of faint, low surface brightness, and ancient MW satellites \citep[M$_* \lesssim 10^5~$M$_\odot$, e.g.,][]{willman2005a,zucker2006,2009MNRAS.397.1748B,drlica-wagner2015,Laevens15,Torrealba16b,2018MNRAS.475.5085T,2019PASJ...71...94H}.  The satellites have stellar populations consistent with the hypothesis that the ionizing radiation from reionization halted or prevented star formation in small ($< 10^9 M_\odot$) halos \citep{1999ApJ...523...54B,2000ApJ...539..517B,2000ApJ...542..535G,2002ApJ...572L..23S,2002MNRAS.333..177B,2008MNRAS.390..920O,2014ApJ...796...91B,2019ApJ...886L...3R}.  Even above this scale, galaxy formation is inefficient, meaning that even relatively large halos have few observable baryons \citep[e.g.,][]{zolotov2012,2016MNRAS.456...85S,2016ApJ...827L..23W,2017arXiv170506286M,2017MNRAS.471.3547F,2018ApJ...863..123B,2018ARA&A..56..435W,2019MNRAS.485.5423D,2020MNRAS.497.1508R,2020arXiv200811207A,2021arXiv210504560G}.  Combined with the selection function for these surveys \citep[e.g.,][]{2008ApJ...688..277T,koposov2009, 2009AJ....137..450W, 2014ApJ...795L..13H,Newton:2017xqg,2020ApJ...893...47D} and the hypothesis of abundance matching to a CDM halo mass function, halos above $\sim$10$^8M_\odot$ must be populated by visible galaxies, and the agreement between the completeness-corrected and predicted satellite luminosity functions is good  \citep{2018MNRAS.473.2060J,2018PhRvL.121u1302K,2019ApJ...873...34N}.  If anything, depending on one's hypothesis for the radial distribution of satellites in the MW, the MW may have (significantly) more galaxies than predicted by CDM models with galaxy formation physics---a \emph{too many satellites} problem \citep{2018PhRvL.121u1302K,2019MNRAS.tmp.1496K}.  This puts severe pressure on numerous novel dark matter models that lead to a truncation in the matter power spectrum \citep{2021PhRvL.126i1101N,2020arXiv201108865N}.

There is more space for new dark matter physics when considering the core-cusp problem, but again baryons also play a significant role in the solution.  Although baryons only make up $\sim$5\% of the universe, simulations show that they play a far-reaching role in the dynamics of dark matter at small scales \citep[e.g.,][]{2005MNRAS.356..107R,2010Natur.463..203G,zolotov2012,2015MNRAS.454.2092O,2016MNRAS.459.2573R, 2017MNRAS.472.2945R, 2019ASSP...56...19B,2019MNRAS.486.4790B,2019MNRAS.483.1314B}.  Diverse density profiles are observed from dwarf- to cluster-scales \citep[e.g.,][]{2001AJ....122.2381M,2011MNRAS.416..322D,2015ApJ...814...26N,2016AJ....152..157L,2017MNRAS.464.2419S,2019MNRAS.484.1401R,2019ApJ...887...94R}.  While baryons bring theoretical predictions closer in line with observations, some suggest that collisionality (i.e. self-interactions) in the dark sector is necessary to explain the full range of observations \citep[e.g.,][]{2014PhRvL.113b1302K,2017PhRvL.119k1102K,2017MNRAS.468.2283C,santos-santos2020}.

A remaining problem is that halo properties predicted by abundance matching do not match those inferred from observed kinematics \citep{2014ApJ...782..115M,2018PhR...761....1B}.  This problem is sometimes called the ``too big to fail'' \citep{2011MNRAS.415L..40B} or ``too big to fail in the field" problem \citep{2015A&A...574A.113P}.  As baryonic tracers probe only the inner $\lesssim$10\% of the virial radius, the inferred dynamical mass within this region must be extrapolated to derive a halo's virial mass.  This puts galaxies of a given stellar mass in a much less massive CDM halo than does traditional abundance matching \citep[e.g.,][]{2019MNRAS.488.3143B}.  Understanding this relationship is important in order to use luminous galaxies to trace halo populations, and to determine which combination of processes are responsible for shaping halo density profiles \citep[see][]{2020ApJS..247...31L}.

In order to find a self-consistent explanation for the abundance of halos, their central densities, and the kinematics of the galaxies they host, it is useful to revisit the missing satellites problem in its original framing \citep{1999ApJ...522...82K, 1999ApJ...524L..19M}.  The key observables of the MW's satellites are their apparent magnitude and internal stellar kinematics.  The original formulation of the missing satellites problem focused on the latter, mapping the line-of-sight stellar velocity dispersion $\sigma^*_{\mathrm{los}}$ of satellites to halo circular velocities using $v_\mathrm{circ} = \sqrt{3} \sigma^*_\mathrm{los}$.  Implied in this relation is a strong prior on the shape of the halo density profile.  The great overabundance of halos at fixed $v_\mathrm{circ}$ in baryon-free CDM simulations relative to the observed $v_\mathrm{circ}$ distribution function was interpreted to mean that a great number of simulated satellites at fixed $v_\mathrm{circ}$ were missing in observations.  Another interpretation, however, is that the mapping between $\sigma^*_{\mathrm{los}}$ and $v_\mathrm{circ}$ is incorrect, because the prior on the density profiles of halos was wrong on account of baryonic feedback or novel dark matter physics \citep{zolotov2012,2018PhR...761....1B}.  Thus, the velocity function version of the missing satellites problem depends both on the abundance of small halos and the matter distribution within them \citep[see][for the stellar velocity dispersion function for massive satellites in MW environments in simulation]{2016ApJ...827L..23W}.    

In this work, we extend our work on completeness corrections of the MW satellite \emph{luminosity} function \citep{2018PhRvL.121u1302K} to the original missing satellites framework of the satellite \emph{velocity} function.  As in our previous work, we perform the completeness correction using several physics-motivated hypotheses for the spatial distribution of satellites throughout the MW halo's volume, and consider uncertainties related to measurement uncertainty and the tidal state of observed satellites (\S \ref{sec:ccVFs}).  We use analytic scaling relations and their intrinsic scatter in order to predict the velocity ($\sigma^*_{\mathrm{los}}$) function for the MW satellites for several baryonic feedback and dark-matter models (\S \ref{sec:theory}). The corrected velocity function matches the predicted CDM velocity function with our most conservative choices for both the completeness correction and the analytic model (\S \ref{sec:results}).  Less conservative choices result in a \emph{too many satellites} problem.  We argue that the ultrafaint dwarf galaxies favor central densities at least as high as predicted by CDM, but that massive classical dwarf galaxies must have cores or shallow cusps.  In \S \ref{sec:discussion}, we show how alternative choices for specific theoretical scaling relations can even further narrow the gap between the theoretical and observed velocity functions.  We summarize our key takeaway points in \S \ref{sec:resolution}.


\begin{figure}  
\centering
\includegraphics[width=0.5\textwidth,clip,trim=0cm 0cm 1cm 1cm]{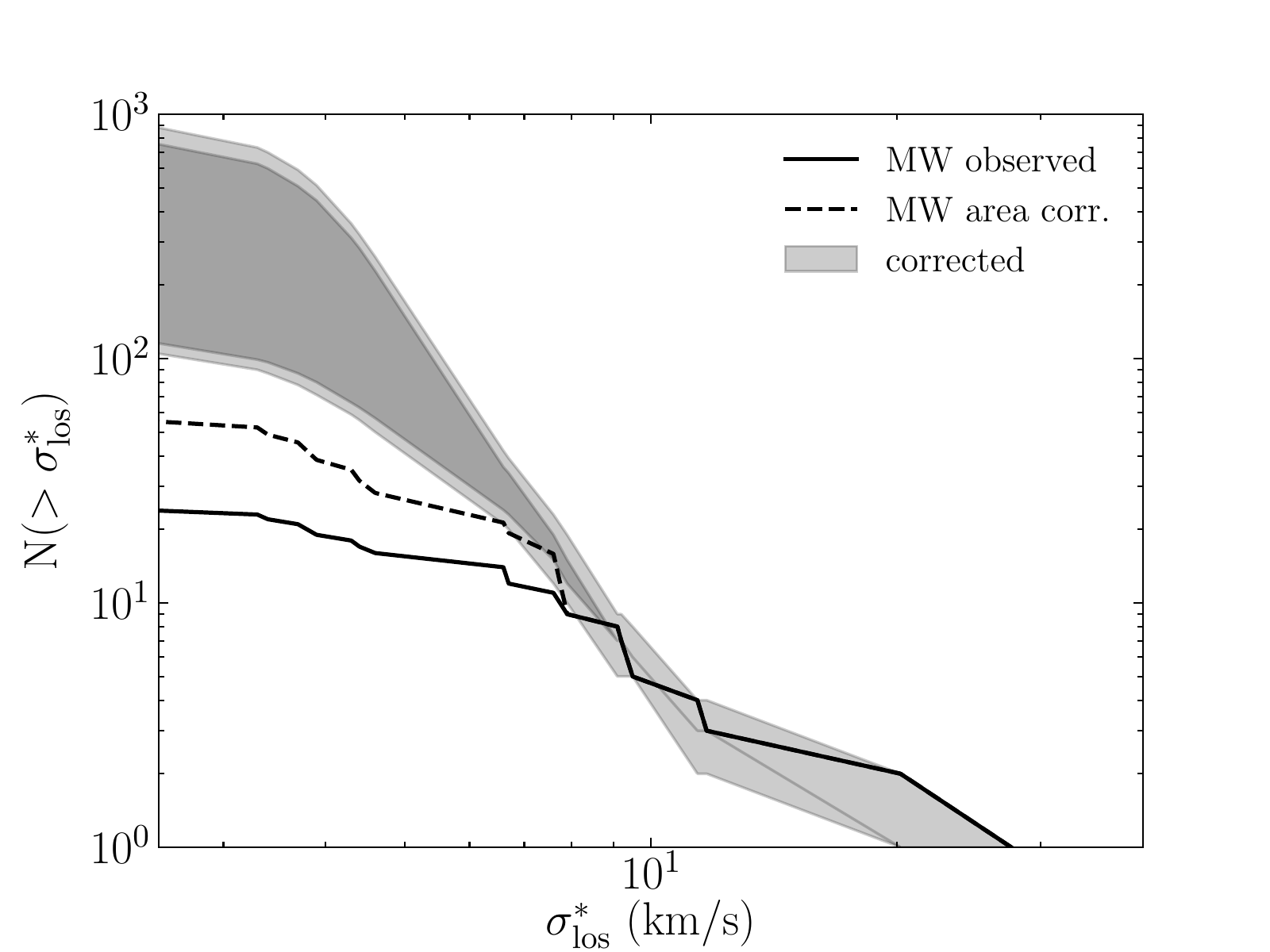}
\caption{The completeness-corrected velocity function (gray band) based on satellites discovered in the SDSS survey.  The dark gray band encompasses the uncertainty due to the unknown radial distribution of satellites.  The bottom edge of the band assumes the distribution mirrors the NFW profile of the underlying MW halo, while the upper edge includes the effect of satellite destruction by the MW's disk.  The light gray bands denotes the measurement uncertainties on $\sigma_{\rm los}^*$ as well as the scatter due to the anisotropy of the MW satellite distribution.  In comparison, the velocity function for the observed classical and SDSS dwarfs is shown with a black line.  Including a simple area correction (i.e. no radial correction) for the SDSS dwarfs would produce the black dashed line.}
\label{fig:vfxns-notheory}
\end{figure}

\section{The Completeness Corrected Velocity Function}
\label{sec:ccVFs}

In this section, we completeness correct the MW's satellite velocity function (VF) following the methodology of \citet{2018PhRvL.121u1302K}, which we previously applied to the MW satellite luminosity function.  In \S\ref{sec:fiducialccvf}, we present our fiducial model for the completeness corrected VF.  In \S\ref{sec:obs_uncert}, we discuss how the uncertainties in the measured stellar velocity dispersions and tidal states of the satellites may influence the completeness correction.

\subsection{Fiducial completeness corrected velocity function}\label{sec:fiducialccvf}

As in \citet{2018PhRvL.121u1302K}, we apply the completeness correction to the SDSS satellite sample \citep{2008ApJ...686..279K,2009AJ....137..450W}.  In brief, the completeness correction is
\begin{equation}
c(\sigma^*_{\rm los}, L) = \frac{\int_{V_\text{vir}} n(\mathbf{r}) ~ d\mathbf{r}}{\int_{V_\text{obs}(L)} n(\mathbf{r}) ~ d\mathbf{r}} \label{eqn:cM}
\end{equation}
where $n(\mathbf{r})$ is the 3D satellite distribution, $V_\text{vir}$ the MW virial volume, and $V_\text{obs}(L)$ the volume over which satellites of luminosity $L$ has been surveyed.  $V_\text{obs}(L)$ depends on both the area covered by the survey as well as the luminosity-dependent distance to which the survey is complete.

We estimate the total number of satellites with a line-of-sight (los) stellar velocity dispersion $\sigma^*_{\rm los}$ by integrating over the observed VF weighted by each satellite's completeness correction,
\begin{equation}
    N(>\sigma^*_{\rm los}) = \int c(\sigma^*_{\rm los},L) ~ \frac{dN_{\rm obs}}{d\sigma^*_{\rm los}} ~ d\sigma^*_{\rm los} ~ = ~ \Sigma^{N_{\rm obs}}_{i=1} c(\sigma^*_{\rm los},L)
\end{equation}
assuming all analogs of a given satellite with luminosity $L$ have roughly the same $\sigma^*_{\rm los}$.  While there is significant scatter in the relationship between $L$ and $\sigma^*_{\rm los}$ \citep{2019ARA&A..57..375S}, the SDSS sample includes many dwarfs of similar $L$ in the regime where the scatter is large, allowing us to sample across this scatter.  As in our earlier work, we assume a spherically symmetric distribution of satellites.

The largest source of uncertainty in the completeness correction is the spatial distribution of satellites $n(r)$.  In Fig. \ref{fig:vfxns-notheory}, the completeness-corrected velocity function is shown as a dark gray band.  The band encompasses the corrections predicted by the range of distributions in the literature.  The lower end is bracketed by the Navarro-Frenk-White \citep[NFW; ][]{1996ApJ...462..563N} distribution, which corresponds to the assumption that the satellites follow the distribution of the MW's dark matter halo.  This results in a high density of satellites in the inner regions of the MW, which requires highly tidally stripped satellites in the MW's central regions to retain a galaxy-like morphology rather than be destroyed or sheared into stream-like debris.  Highly resolved numerical studies of the tidal stripping of dwarfs indicate that cuspy dwarfs likely survive under the tidal forces of the MW \citep{2019PhRvD.100f3505D,2020MNRAS.491.4591E}.

In contrast, if tidal stripping destroys a significant fraction of satellites in the inner MW, the resultant satellite distribution produces the upper end of the dark gray shaded region.  In detail, distributions derived by selecting a subset of subhalos from dark matter only (DMO) simulations to host galaxies \citep[e.g. those that were most massive in the past, had the earliest infall, or are reionization fossils; ][]{2014ApJ...795L..13H}, which lie in the middle of the gray band, were modified to reflect the destruction of satellites due to the MW's baryonic disk following \citet[][see also \cite{2010ApJ...709.1138D, 2013ApJ...765...22B,2020MNRAS.492.5780R,2020MNRAS.491.1471S}]{2017MNRAS.471.1709G}.

The light gray bands denote the 10 and 90\% percentile satellite counts from MCMC realizations of the completeness-corrected satellite population based on the uncertainties in the measurements of $\sigma_{\rm los}^*$.  Further, a 30\% scatter as been added to account for anisotropy \citep{2008ApJ...688..277T}.  For simplicity, we assume that uncertainties in measured $\sigma_{\rm los}^*$ are Gaussian.  The effect of other sources of uncertainty, such as the mass of the MW, are not shown in the uncertainty to the completeness corrections but are explored in \S\ref{sec:discussion}.

We adopted the measured line-of-sight (los) stellar velocity dispersions $\sigma^*_{\rm los}$ and the associated measurement uncertainties for MW dwarfs cataloged in \citet{2012AJ....144....4M}, with one exception:  Bootes II.  \citet{2012AJ....144....4M} lists Bootes II's as $\sigma^*_{\rm los} = 10.5 \pm 7.4$ km/s based on Gemini/GMOS measurements of five stars \citep{2009ApJ...690..453K}.  However, we adopt $\sigma^*_{\rm los} = 4.4 \pm 1$ km/s based on 23 confirmed members observed with Keck/DEMIOS and reduced using the same methods as in \citet{2009ApJ...692.1464G} (Geha, private communication).  The Keck/DEIMOS velocity dispersion for Bootes II is consistent with the measured velocity dispersion of other dwarf galaxies of comparable stellar mass.


\subsection{Observational uncertainties}
\label{sec:obs_uncert}

\begin{figure}
    \includegraphics[width=0.5\textwidth]{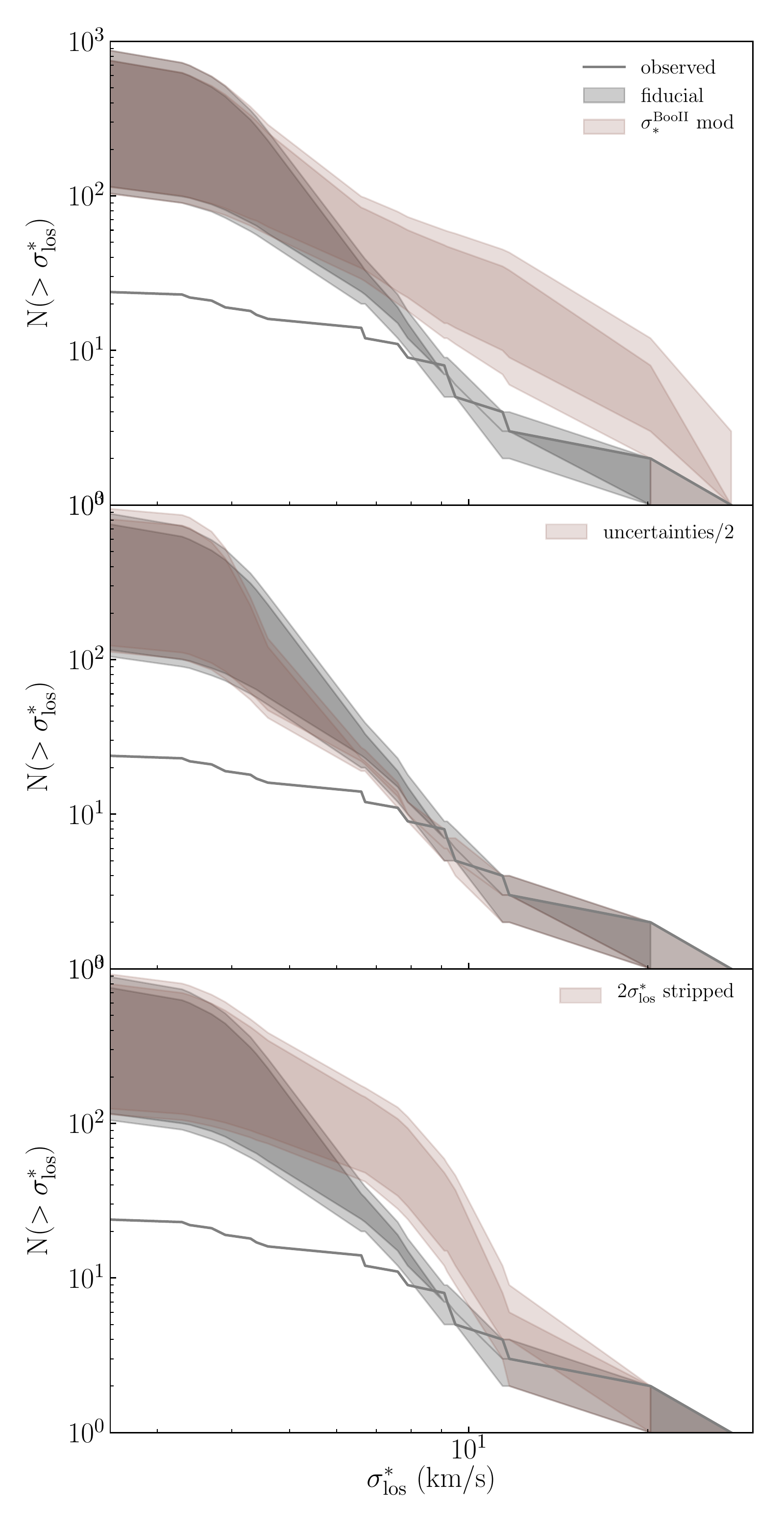}
    \caption{Same as Fig. \ref{fig:vfxns-notheory}, but (\emph{top}) without the revised measurement of Boo II's $\sigma_{\rm los}^*$ by Geha (private communication), (\emph{middle}) with measurement uncertainties reduced by a factor of two, or (\emph{bottom}) with the $\sigma_{\rm los}^*$ higher by a factor of 2 for the unobserved analogs of satellites with evidence of tidal stripping.}  
    \label{fig:vfxn_obs_uncert}
\end{figure}

Here, we consider sources of uncertainty in our completeness correction on account of the measurement of the the velocity dispersions of stars in the satellites, and the tidal state of the galaxies.  

First, we consider measurement uncertainties.  While measurements of $\sigma_{\rm los}^*$ for the classical satellites are based on hundreds to thousands of member stars, far fewer stars are available for ultrafaints satellites, introducing significant systematic measurement uncertainties.  Compounding this issue, there is evidence that binary stars may make up a higher fraction of low-metallicity populations, which may artificially drive up velocity dispersions \citep{2019ApJ...875...61M}, although their typical impact may be small \citep[e.g.][]{2019ARA&A..57..375S}.   Both of these were identified as issues with Bootes II's original $\sigma_{\rm los}^*$ measurement of 10.5 $\pm$ 7.4 km/s based on 5 member stars \citep[][ also the value tabulated in \citet{2012AJ....144....4M}]{2009ApJ...690..453K}.  In our analysis, we used an updated measurement derived using Keck/DEMIOS spectroscopy of 23 members and reduced using the methods in \citet{2009ApJ...692.1464G}, which gave a value of 4.4 $\pm$ 1.0 km/s (Geha, private communication).

The completeness-corrected VF changes significantly depending on which value is used.  To illustrate, in the top panel of Fig. \ref{fig:vfxn_obs_uncert}, we compare the corrected VFs inferred utilizing the earlier (brown) and updated (fiducial; gray) measurement.  The corrected VF becomes significantly less steep and shifts upwards particularly at the high velocity end.  The kink at $\sim$10 km/s disappears, and the VF becomes scale-free.  This affects our theoretical interpretation of the VF, as described in \S\ref{sec:discussion}.

There is evidence that Bootes II is an outlier in the outsize impact binaries had on its $\sigma_{\rm los}^*$ measurement.  However, there is still significant uncertainty in the velocity measurements of faint dwarfs, and the need for improved measurements still remains.  Segue 2 and Willman 1, ultrafaints measured to have $\sigma_{\rm los}^*$ = 3.4$^{+2.5}_{-1.2}$ km/s \citep{2009MNRAS.397.1748B} and 4.3$^{+2.3}_{-1.3}$ km/s \citep{2007MNRAS.380..281M}, respectively, were both subsequently reanalyzed, which produced only upper limits on $\sigma_{\rm los}^*$ \citep{2013ApJ...770...16K, 2011AJ....142..128W}.  Willman 1 appears to have a high velocity dispersion in its outskirts, suggesting that tidal effects may be at play.  For both dwarfs, we chose to use their original values, but lower values could lower the corrected velocity functions. Indeed, the broader sample of $\sigma_{\rm los}^*$ reassessed with new Keck/DEIMOS spectroscopy indicates that the updated measurements tend to be lower than previously published values (Geha, private communication). However, the first results from a homogeneous reanalysis of VLT/GIRAFFE data indicate that the velocity dispersion can move up or down, and is highly sensitive to stellar membership criteria as well as to binary stars \citep{2021arXiv210100013J}.  Thus, controlling systematic uncertainties is important to defining the overall shape of the VF.

Controlling statistical uncertainties is also key.  To demonstrate the power of reduced statistical measurement uncertainties, we show the corrected VF with uncertainties on $\sigma_{\rm los}^*$ reduced by a factor of two in the middle panel of Fig. \ref{fig:vfxn_obs_uncert}  (brown; adopting our fiducial $\sigma^*_{\rm los, BooII}$ = 4.4 km/s), assuming the measured values of $\sigma_{\rm los}^*$ are correct.  Again, the fiducial completeness correction with the current uncertainties are shown in gray.  Smaller uncertainties lowers the slope of the corrected VF and sharpens the break in the VF power-law relation near $\sigma_{\rm los}^* = 10$ km/s. These features would be even more pronounced if the measured values of $\sigma_{\rm los}^*$ themselves were lower.

\subsubsection{Tidal stripping}

While $\sigma^*_{\rm los}$ does not change much if a subhalo is stripped down to 10\% of its initial mass---about 5-15\%, producing nearly imperceptible changes to the velocity function---severe stripping down to 1\% of a subhalo's original mass can reduce $\sigma^*_{\rm los}$ by as much as 40-60\%, with the exact value dependent on the mass of the subhalo and its density profile \citep[see the discussion of Fig.~\ref{fig:menc} in \S\ref{sec:theory};  also][]{2018MNRAS.478.3879S}.  This implies that the assumption we made that the likely less stripped (on account of their distance from the host), unobserved analogs have the same $\sigma^*_{\rm los}$ as their observed counterparts may fail.  The analogs might instead be assigned higher $\sigma^*_{\rm los}$ if the nearby observed ultrafaint dwarf galaxies are highly stripped.

Identifying which dwarfs have undergone severe stripping is non-trivial, and the degree to which individual Milky Way dwarfs have been tidally stripped is not well constrained.  As galaxies are deeply embedded within their dark matter halo, stellar mass loss does not begin until $>$90\% of the halo has been stripped \citep{2008ApJ...673..226P, 2015MNRAS.449L..46E}.  Galaxies that exhibit signs of tidal perturbations are already significantly stripped.  A few dwarfs do exhibit such perturbations.  The clearest example among the ultrafaint dwarfs is Hercules, which exhibits an elongated morphology, a velocity gradient, and is surrounded by tidal debris \citep{2007ApJ...668L..43C,2009ApJ...704..898S,2012MNRAS.425L.101D,2015ApJ...804..134R,2017ApJ...834..112K,2018ApJ...852...44G,2020ApJ...902..106M}.  Its orbit lends support for this scenario \citep{2018ApJ...863...89S}.  The case for other potentially heavily stripped dwarfs is perhaps less clear.
Leo V exhibits several hints that suggest it has been tidally disturbed:  an elongated morphology \citep{2008ApJ...686L..83B}, a stellar distribution that extends as far as 10 $R_{\rm eff}$ and is more metal-rich than other dwarfs of its luminosity \citep{2009ApJ...694L.144W}, a stellar stream-like overdensity \citep{2012ApJ...756...79S} though \citet{2019ApJ...885...53M} find they are likely field stars, and a velocity gradient and significant [Fe/H] spread \citep{2017MNRAS.467..573C}.  However, its inferred pericenter of $\sim$170 kpc \citep{2020MNRAS.494.5178F} suggests it has not experienced significant tidal perturbations from the MW.
Segue 2 is also significantly more metal-rich than other dwarfs of similar luminosity, suggesting that it may have been a classical dwarf as massive as Ursa Major but experienced heavy stripping to become an ultrafaint dwarf \citep{2013ApJ...770...16K}.
Willman 1's unusual kinematics and large metallicity spread suggest it may also have undergone significant stripping, although given the uncertainty in its velocity dispersion, it remains possible it may be a globular cluster \citep{2011AJ....142..128W}. 

Given these uncertainties, we present a thought experiment:  if Hercules, Leo V, Segue 2, and Willman 1 have been stripped to $\sim$1\% of their infall mass, how would this affect the corrected VF?  We assign the unobserved analogs of these galaxies inferred from our completeness corrections a $\sigma^*_{\rm los}$ a factor of two higher than their observed counterpart.  The resulting completeness-corrected VF is shown in the bottom panel of Fig. \ref{fig:vfxn_obs_uncert}.  The VF shifts dramatically at $\sim$10 km/s.  As we show in \S\ref{sec:discussion}, this has significant implications for our theoretical interpretation of the satellites. A better understanding of the extent of tidal stripping of the faint dwarfs that dominate the completeness corrections is essential for a robust estimate of the corrected VF.

\vskip 0.2cm

For the remainder of this work, we will compare theoretical models to the fiducial completeness-corrected VF described in Fig.~\ref{fig:vfxns} in Sec.~\ref{sec:fiducialccvf}.  If Boo II's velocity dispersion is as large as indicated in \citet[][with the caveat that the uncertainty on that measurement is large]{2009ApJ...690..453K}, it would be a significant outlier from ultrafaints of comparable luminosity.  The revised velocity dispersion measurement by Geha is more in line with other satellites, thus we adopt it for our completeness-corrected VF. Furthermore, we consider that the probability that a large fraction of nearby ultrafaints are stripped enough to alter the VF significantly from our fiducial correction to be small.  If the satellites were so stripped, though, the completeness-corrected VF would even more strongly indicate a ``too many satellites problem", as discussed in Sec.~\ref{sec:results}.


\section{Theoretical Predictions for the Velocity Function}
\label{sec:theory}

To evaluate the small-scale implications of the completeness-corrected VF, we here present our framework and modeling choices for theoretical predictions of the MW's VF.  In order to explore the impact of specific physical models for halos, galaxies, and the connection between the two for CDM and other dark matter models, we adopt an analytic approach based on scaling relations, as in our previous work \citep{2018PhRvL.121u1302K}.  Unlike our previous work, we fully include scatter in the scaling relations.  Because high-resolution simulations that include baryons are extremely costly and few in number (especially if restricted to those that can resolve ultrafaint dwarf galaxies, M$_* < 10^5$~M$_\odot$), our analytic approach enables us to efficiently test well-motivated scaling relations.  In \S \ref{sec:discussion}, we systematically consider deviations from our fiducial choices in order to explore the range of plausible VFs for a wide range of variations to our standard model.  In future work, all sources of uncertainty can be considered and marginalized over.

We translate a theoretical halo mass function into a theoretical VF.  We adopt the subhalo mass function from the ELVIS simulations \citep{2014MNRAS.438.2578G} of the Local Group, but with a normalization that is lower by 20\% to reflect the suppression of the halo mass function by baryons relative to the ELVIS dark matter only (DMO) simulations \citep{2017MNRAS.471.1709G}.  This is consistent with other work \citep{2020MNRAS.493.1268B, 2020MNRAS.492.5780R, 2020MNRAS.491.1471S}.  Note that we employ the \emph{infall} mass function, i.e. the masses of the satellites as they enter the MW's virial radius.  Infall halo masses are defined such that the mean over-density within the virial radius $R_{\rm 200}$ relative to the critical density is $\Delta$ = 200.  We assume a MW mass of 10$^{12}$ M$_\odot$ \citep{2020SCPMA..6309801W}.

Host halos of a given mass exhibit scatter in the number of subhalos that inhabit them due to differences in the host halos' accretion histories \citep[e.g.][]{2019MNRAS.486.4545F}.  We generate 100 realizations of the MW subhalo population, sampling the halo-to-halo scatter in the number of subhalos measured by \citet{2010MNRAS.406..896B} in the Millennium-II Simulation, who found the scatter to have both a Poissonian and an intrinsic component,
\begin{equation}
\sigma^2 = \sigma^2_P + \sigma^2_I,
\end{equation}
where $\sigma_P^2$ = $\langle N(>\mu)\rangle$ and is the Poisson component, $\sigma_I^2$ = 0.18 $\langle N(>\mu)\rangle$ is the intrinsic component, and $\mu$ = $M_{\rm subhalo}/M_{\rm MW}$.  For each realization of the MW satellites, we assign each subhalo a mass selected from the ELVIS subhalo mass function.

Not all subhalos host galaxies.  Ionizing UV radiation from young stars heated the gas within galaxies during the epoch of reionization, reducing the availability of dense, cold gas from which stars could form \citep{2000ApJ...539..517B, 2000ApJ...542..535G, 2014ApJ...796...91B, 2014MNRAS.442.1396W, 2019ApJ...886L...3R}.  Only halos that had reached masses above $\sim$10$^{7-8}$ M$_\odot$ by reionization would have had gas with the ability to cool and continue to form stars \citep{Koh2018,2019ApJ...886L...3R}.  To select for halos with galaxies, we used $f_\text{lum}(M)$, the fraction of halos with an infall mass $M$ that contain a galaxy, from \citet{2017MNRAS.471.4894D}.  This relation assumes a reionization redshift $z_{\rm re}$ = 9.3 and places galaxies in halos with $v_{\rm max} > 9.5$ km/s by reionization or halos with a present-day $v_{\rm max} > 23.5$ km/s.  Our pre-reionization $v_{\rm max}$ threshold lies somewhat below the atomic cooling limit \citep{Koh2018}, and the post-reionization threshold is consistent with but lower than that of \citet{2008MNRAS.390..920O}.  It is a conservative choice---higher values of $v_{\rm max}$ reduce the number of luminous satellites. This process produces the observable satellite mass function.

\begin{figure}
    \centering
    \includegraphics[width=0.5\textwidth,clip,trim=0cm 0.5cm 0cm 0cm]{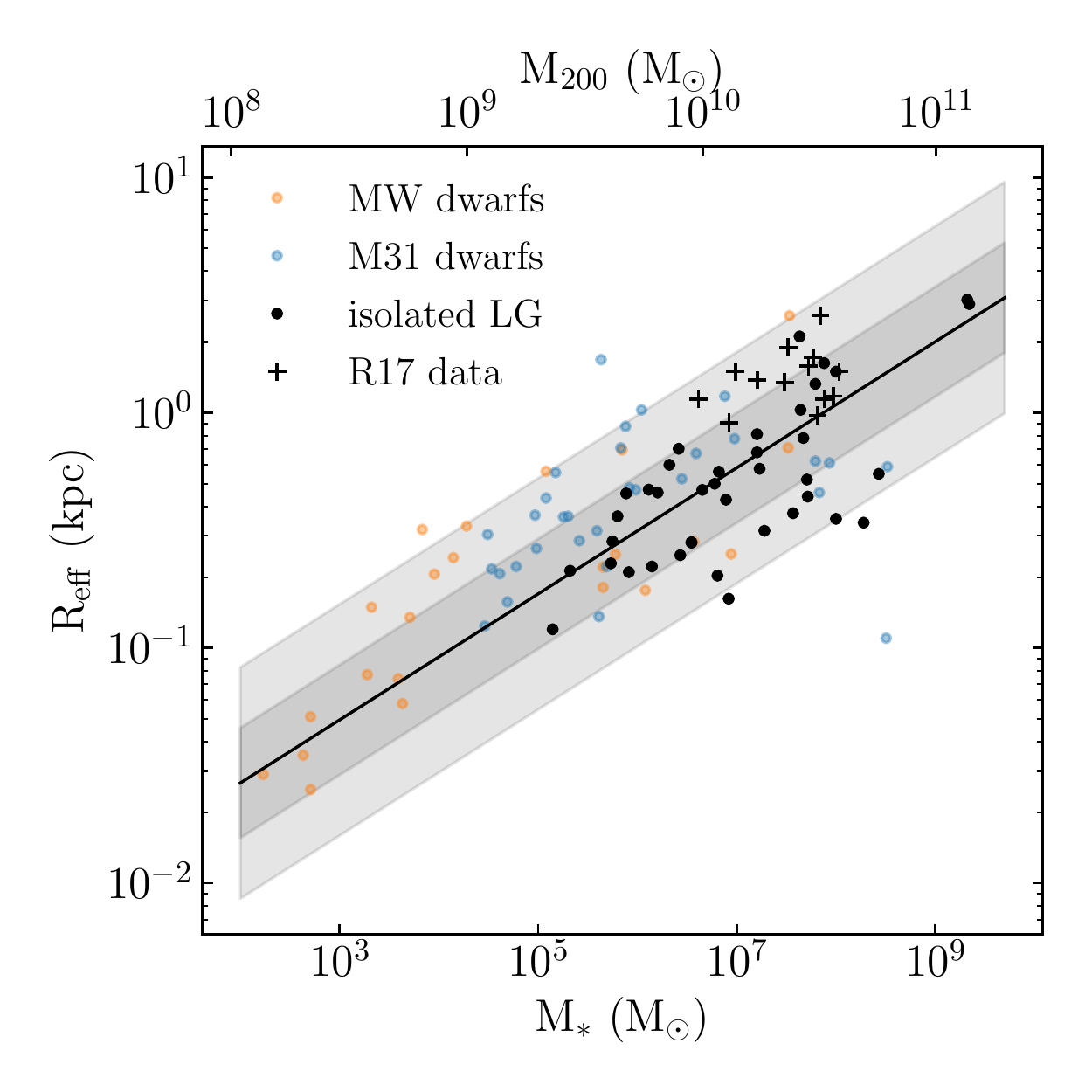}
    \caption{Stellar mass-size relation for dwarf galaxies in the Local Group.  The projected 2D half-light radius is shown.  Data for dwarfs plotted as a circle were taken from \citet{2012AJ....144....4M}, while those plotted with a cross are from \citet{2017MNRAS.467.2019R}.  A fit to isolated dwarfs from both works is shown via the black line.  The 1- and 2-sigma bands are shown via the gray and light gray bands, respectively.  The corresponding dark matter halo mass M$_{200}$ is derived using the \citet{2013MNRAS.428.3121M} stellar-mass--halo-mass relation.}
    \label{fig:mvir_dependence}
\end{figure}

To compare with our completeness-corrected VFs, we then map halo masses into line-of-sight stellar velocity dispersions $\sigma^*_{\rm los}$.  \citet{2010MNRAS.406.1220W}'s mass estimator can be rearranged to give
\begin{equation}\label{eq:wolf}
\langle \sigma^{*,2}_{\rm los} \rangle = \frac{G}{4} \frac{M_{1/2}}{R_{\rm eff}},
\end{equation}
where $R_{\rm eff}$ is the 2D projected half-light radius and $M_{1/2}$ is the mass enclosed by the 3D half light radius $r_{1/2}$.  We convert between the 2D and 3D half-light radii via $r_{1/2} = R_{\rm eff}/0.75$, an approximation accurate for a wide variety of stellar light distributions \citep{2010MNRAS.406.1220W}. We also considered the mass estimator from \citet{2018MNRAS.481.5073E}, which produce small differences in $\sigma^*_{\rm los}$, ranging from $<$5\% for NFW and $<$20\% for SIDM profiles, although this increases (decreases) slightly with increasing tidal stripping for NFW (SIDM) profiles.  The resultant changes to the velocity function are small and do not affect our conclusions.

For the 2D half-light radius $R_{\rm eff}$, we fit an empirical relation between $R_{\rm eff}$ and the stellar mass $M_*$ of isolated dwarfs from \citet{2012AJ....144....4M} and \citet{2017MNRAS.467.2019R}, which gives
\begin{equation}
    \log_{10} R_{\rm eff} = 0.239 \log_{10} M_* - 1.68
\end{equation} 
with a 1$\sigma$ scatter of 0.234 dex.  This relation is shown in Fig. \ref{fig:mvir_dependence}.  This mass-size relation is similar to that found by \citet{2018ApJ...856...69D} for observed satellites in the Local Group, as well as in simulations \citep[e.g.][]{2019MNRAS.488.4801J}.  We sample the scatter to generate a $R_{\rm eff}$ for each subhalo in our satellite realizations.  We derive the stellar mass $M_*$ from the infall halo mass via an extrapolation of the stellar-mass--halo-mass (SMHM) relation by \citet{2013MNRAS.428.3121M}, which matches results from hydrodynamic simulations \citep{2015MNRAS.453.1305W,2017arXiv170506286M,2019ApJ...874...40M,2019MNRAS.483.1314B,2020arXiv200811207A}, sampling a $M_*$ assuming a 1$\sigma$ scatter of 0.15 dex.  We discuss the effect of adopting alternative SMHM relations in \S\ref{sec:discussion}.

\begin{figure*}
    \centering
    \centerline{
    \includegraphics[width=0.5\textwidth,clip,trim=0cm 0cm 1.2cm 0cm]{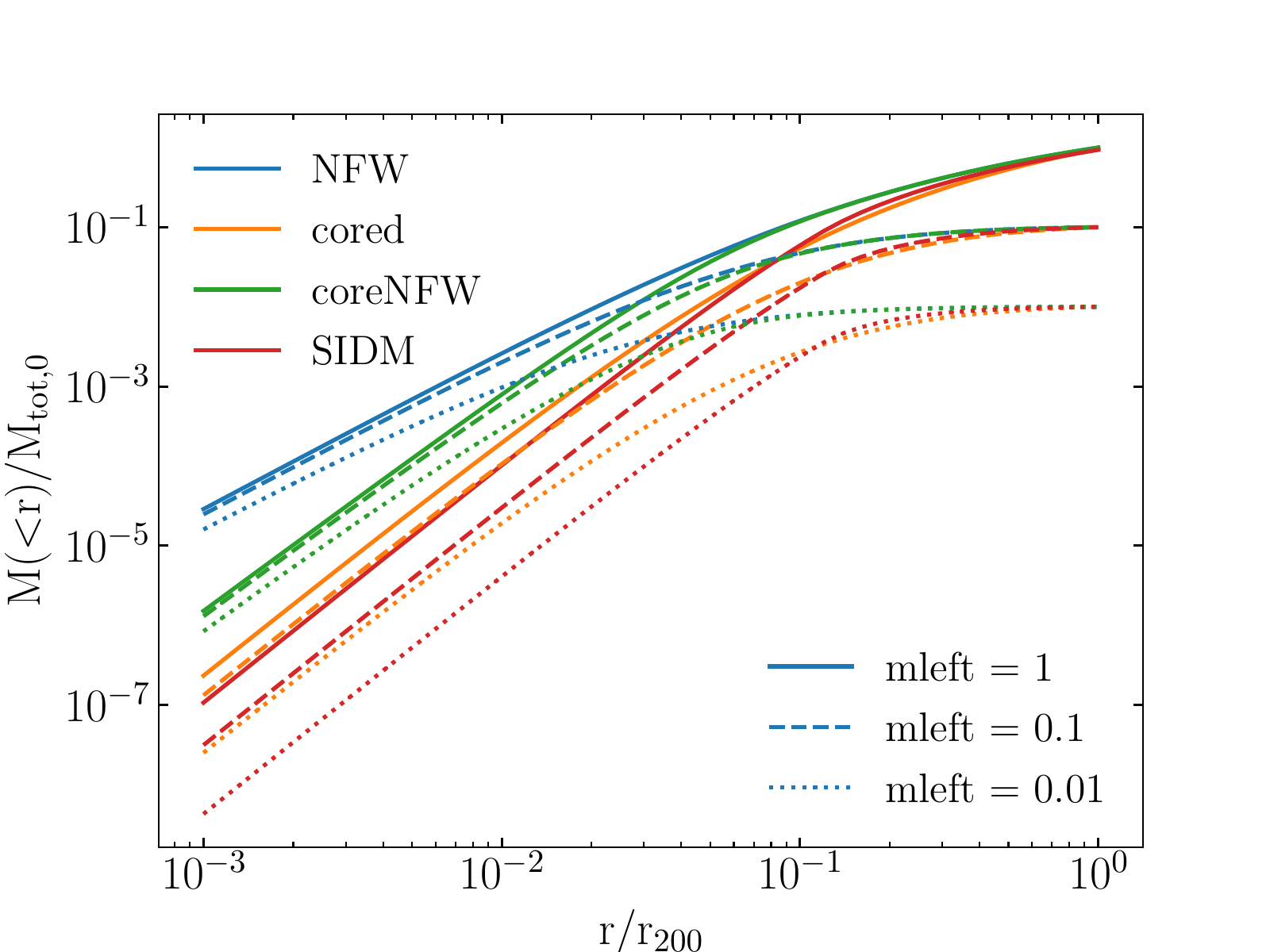}
    \includegraphics[width=0.5\textwidth,clip,trim=0cm 0cm 1.2cm 0cm]{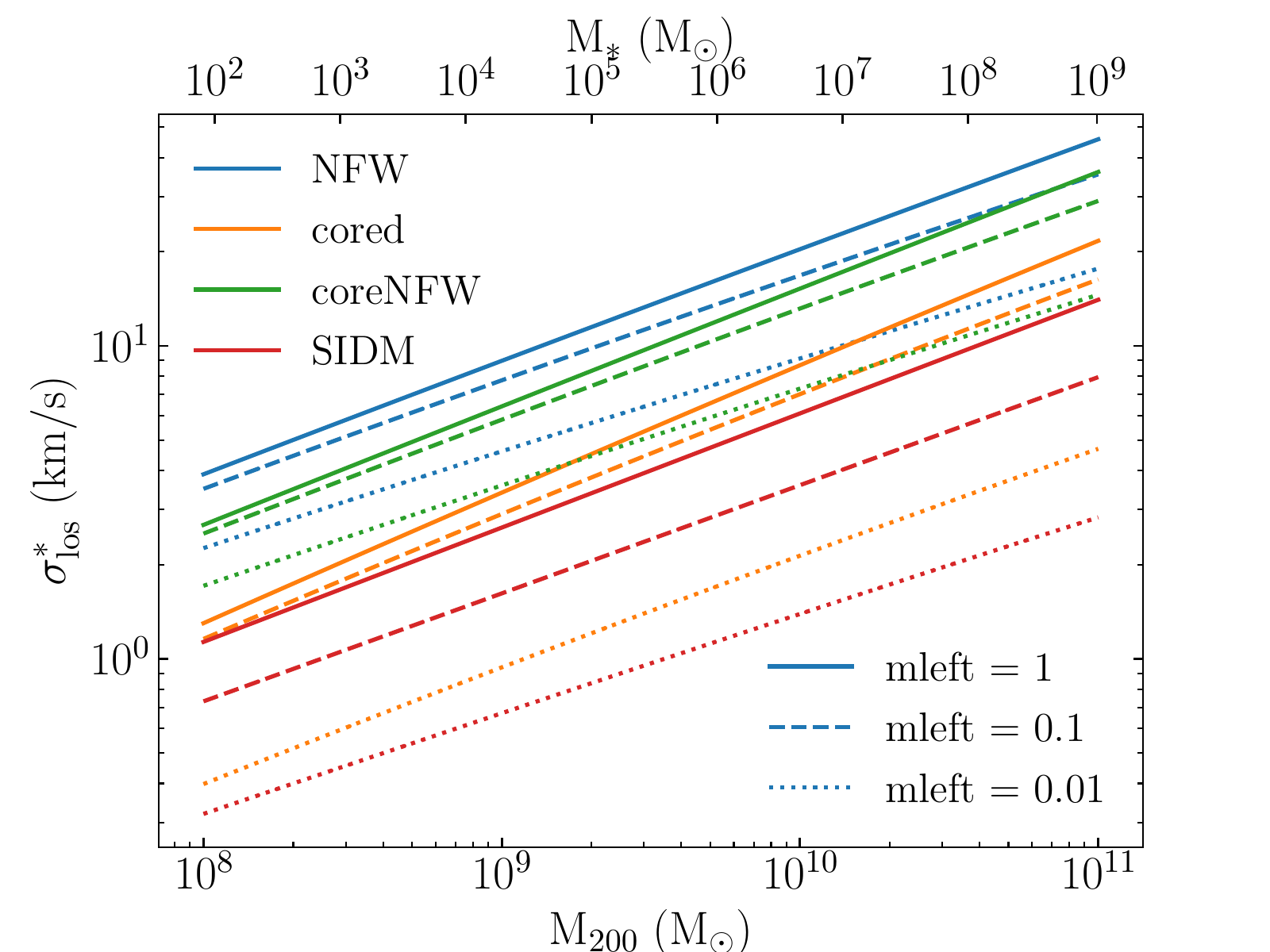}} 
    \caption{Mass enclosed (left) for a 10$^9$ M$_\odot$ halo and the translation between $M_{\rm vir}$ at infall and $\sigma^*_{\rm los}$ (right) for NFW, core ($\alpha\beta\gamma$ = 1,3,0), \citet{2016MNRAS.459.2573R}'s coreNFW, and SIDM ($\sigma/m_\chi = 1$ cm$^2$/g) profiles.  The line styles denote different levels of tidal stripping.  We have used our fiducial choice of $z_{\rm in}$ = 1 as a representative infall time for the MW satellites.  The stellar contribution to the density profiles nor the velocity dispersions are not included in either panel.}
    \label{fig:menc}
\end{figure*}

Given a density profile for the halos, we can derive the mass within the 3D half-light radius $M_{1/2}$ by integrating the profile to $r_{1/2}$.  The profiles we consider are discussed in \S\ref{sec:density_profiles} and shown in Fig.~\ref{fig:menc}.  The halos of the MW satellites are tidally stripped by the MW, even if their resident galaxies exhibit no visible tidal perturbations.  Stripping modifies halo density profiles, which we discuss in \S\ref{sec:strip}.

\begin{figure}
    \centering
    \includegraphics[width=0.5\textwidth]{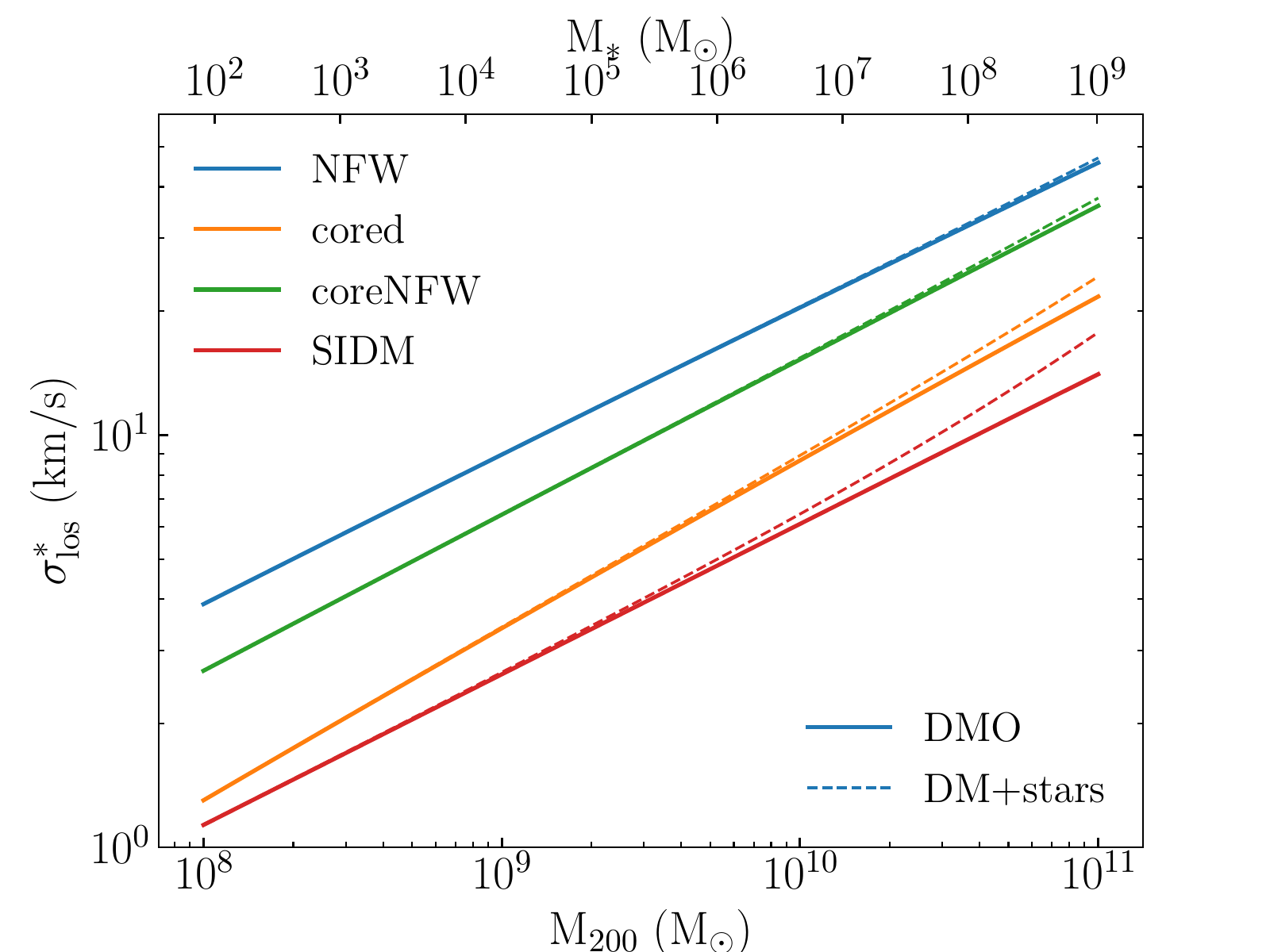}
    \caption{The change in the translation between the dark matter virial mass and $\sigma^*_{\rm los}$ when the contribution to the potential by the stars is added, assuming no tidal stripping and the SMHM relation of \citet{2013MNRAS.428.3121M}. Note that the stellar contribution becomes important above 10$^{10}$ M$_\odot$ in DM virial mass or 10$^7$ M$_\odot$ in stellar mass---the stellar mass of Fornax.}
    \label{fig:sigLOS_with_stars}
\end{figure}

Dwarfs with masses at least that of Fornax ($M_* \sim 10^7$ M$_\odot$, $M_{200} \sim 10^{10}$ M$_\odot$) have baryonic components that contribute significantly to $M_{1/2}$.  To account for this, we computed the stellar mass $M_*$ for each halo via \citet{2013MNRAS.428.3121M}'s SMHM relation and added half this value to the dark matter mass within that radius, assuming half the stellar mass lies within the half-light radius.  Fig.~\ref{fig:sigLOS_with_stars} shows the change in $\sigma^*_{\rm los}$ between the (dark matter) virial mass and that with the stellar mass included.  One can see that the stellar contribution becomes significant for $M_{200} \gtrsim 10^{10}$ M$_\odot$ or equivalently, $M_* \gtrsim 10^7$ M$_\odot$.


\subsection{Density profiles}
\label{sec:density_profiles}

The dark matter density profile(s) of satellite halos has been a matter of significant debate, with claims of universal cuspiness \citep[i.e. a steep inner slope $\alpha \lesssim -1$ if $\rho(r) \propto r^\alpha$ for small $r$,][]{1996ApJ...462..563N}, to large cores ($\alpha \simeq 0$), and others claiming a mix of both.  At dwarf scales, the classic cusp-core problem has given way to a ``diversity'' conundrum---dwarf galaxies show a strong diversity in their central slopes \citep{2005ApJ...621..757S,2008ApJ...676..920K,2015MNRAS.452.3650O,2016MNRAS.462.3628R,2019ApJ...887...94R,2019PhRvX...9c1020R,santos-santos2020,2020ApJ...904...45H}.  To capture this diversity, we model a range of density profiles.

We consider two $\alpha\beta\gamma$ broken power law profiles \citep{1990ApJ...356..359H, 1996MNRAS.278..488Z}: the classic cusped NFW ($\alpha,\beta,\gamma$ = 1,3,1) profile predicted in dark-matter-only CDM simulations, and cored ($\alpha,\beta,\gamma$ = 1,3,0).  While this cored profile has been used in a number of works to study the survival of satellites with cores vs. cusps \citep[e.g.][]{2010MNRAS.406.1290P, 2017MNRAS.465L..59E}, it produces unrealistically large cores relative to those produced by baryonic feedback, as we show below.  However, we include it for completeness.  For both models, we characterize the break in the power law with the halo concentration parameter $c_{200}$.  We use \citet{2019ApJ...871..168D}'s mass-concentration relation at the median infall time of the MW's subhalos, $z_{\rm in} \simeq$ 1 \citep{2012MNRAS.425..231R, 2017MNRAS.471.4894D, 2019arXiv190604180F}, sampling the 0.16 dex 1$\sigma$ scatter in this relation.

For a more realistic cored density profile, such as those produced by baryonic feedback in CDM, we adopt the coreNFW model by \citet{2016MNRAS.459.2573R}.  This model assigns core sizes based on the length of time a dwarf has formed stars, $t_{\rm SF}$.  For simplicity, we assume all satellites stop forming stars at infall.

Finally, we consider cored profiles predicted in self-interacting dark matter (SIDM) models, which have been proposed as a solution to CDM's small-scale issues from the dark sector \citep[see, e.g.][for a review]{2018PhR...730....1T}.  Our model for the SIDM halo density profile is derived from \citet{2013MNRAS.430...81R}, which assumes a constant, velocity-independent self-interaction cross section $\sigma_{\rm SI}/m_\chi$.  In brief, we identify the radius at which the per-particle scattering rate $\Gamma(r)$ is one given the age of the halo, $t_{\rm age}$.  This radius, $r_1$, is given by
\begin{eqnarray}
    \label{eq:sidm_density-rate_of_int}
\rho_s ~ g(r_1/r_s) ~ v_{\rm rms} ~ \frac{4 ~ t_{\rm age}}{\sqrt{3\pi}} ~ \frac{\sigma_{\rm SI}}{m_\chi} = 1\nonumber \;, \hbox{ or}\\
\alpha ~ \frac{v_{\rm max}^3}{G_{\rm N}~R_{\rm max}^2} ~ g(r_1/r_s) ~ t_{\rm age} ~ \frac{\sigma_{\rm SI}}{m_\chi} = 1.
\end{eqnarray}
The parameters $\rho_s$ and $r_s$ are the NFW scale density and radius, and $g(x)=x^{-1}(1+x)^{-2}$ is the scale-free form of the NFW profile, where $x = r/r_s$.  The particle velocity dispersion $v_{\rm rms}\simeq 1.1~v_{\rm max}$, where $v_{\rm max}$ is the maximum circular velocity of the halo and $r_{\rm max}$ the radius at which it occurs.  The parameter $\alpha$ is a scale factor, which \citet{2013MNRAS.430...81R} found to be 2.5 on cluster scales.  At dwarf galaxy scales, we find $\alpha \approx 1.5$ based on fits to the Merry and Pippin halos in \citet[][$M_{200} \sim 10^{10} M_\odot$]{2015MNRAS.453...29E} and the lowest-mass halos of \citet[][$M_{200} \sim 10^{12} M_\odot$]{2013MNRAS.430...81R}.

Pre-infall, the halo is NFW at $r > r_1$ , but follows a Burkert profile inside $r_1$,
\begin{equation}
\rho_B(r) = \frac{\rho_0}{(1+r/r_B)(1+r^2/r_B^2)}, \label{eq:burkert}
\end{equation}
where $\rho_0$ is the normalization of the profile and $r_B$ is the Burkert core radius.  For dwarf-scale halos, $r_B \approx 1.3 r_1$ based on fits to the simulations mentioned above.  We do not consider the effect of baryons in shaping the core, which should be a good approximation for all but the most massive MW satellites \citep[the Clouds,  Sagittarius, and Fornax;][]{2017MNRAS.472.2945R}.  We note that the shape of the cored $\alpha\beta\gamma$ density profile is similar to that of SIDM with an elastic isotropic cross section of $\sigma_{\rm SI}/m_\chi = 1\hbox{ cm}^2/\hbox{g}$.

The enclosed mass profiles and the associated $\sigma_{\rm los}^*$ for each of the density profiles discussed above are shown in Fig. \ref{fig:menc}, including the effects of tidal stripping, which we discuss next.


\subsection{Tidal stripping}
\label{sec:strip}

When a halo enters the neighborhood of a more massive host, tidal forces strip material from it, causing mass loss, orbital decay, and with advanced tidal evolution, the destruction of the subhalo and its resident galaxy.  Although tides mainly operate outside-in, tidally-induced mass loss depletes mass at all radii within a subhalo.  Density profiles are further affected by tidal shocking and tidal heating \citep[e.g.][]{1999ApJ...514..109G,2006MNRAS.366..429R,2016MNRAS.461..710D,2019PhRvD.100f3505D,2020MNRAS.494..378D}.  The mass within $R_{\rm eff}$, and hence $\sigma_{\rm los}^*$, can be affected.  Here, we describe our model for tidal stripping, treating the dark and luminous matter separately.  It is important to note that our goal is to estimate the effects of tidal stripping for a typical satellite of the MW.  While some satellites of the MW may be stripped of more than $\sim 90\%$ of their dark matter, the typical satellite will experience significantly less stripping \citep[see, for example the $M_{\rm peak}$ subhalo mass function vs. the present day-mass halo mass function in][]{2017MNRAS.471.1709G}.  In Sec.~\ref{sec:results}, we discuss the implications of tidal stripping evolution on the interpretation of the VF.

\subsubsection{Evolution of the dark matter halo}

We model the effects of tidal stripping on subhalos according to fitting functions derived in past simulation work.  \citet{2010MNRAS.406.1290P} found that the \emph{structure} of subhalos with $\alpha\beta\gamma$ density profiles largely remains unchanged under tidal stripping by its host halo except at its outskirts, where the slope of the density profile steepens to $\gamma = 5$ \citep[see also][]{2019MNRAS.490.2091G}.  Further, the normalization of the density profile drops slightly.  We adopt $\gamma = 5$ and adjust the normalization of stripped halos according to the procedure detailed in Appendix A3 of \citet{2010MNRAS.406.1290P} before integrating to calculate $M_{1/2}$.  We note that \citet{2010MNRAS.406.1290P} adopted $\Delta \sim$ 100, in contrast to our choice of $\Delta$ = 200.  We have made the relevant conversions between the two mass definitions before applying their methodology. We further note that \citet{2010MNRAS.406.1290P}'s prescription predicts that mass within a small fixed radius $\lesssim$ R$_{200}$/100 to initially increase slightly (less than 10\%) before falling, counter to the expectation that stripping always reduces the mass at all radii.  However, this rise does not occur for the mass within the stripping-dependent half-light radius---the relevant quantity for this work---as shown in Appendix \ref{appendix:notes_on_P10}. 

A similar study of how the coreNFW profile changes under tidal stripping has not been done. However, the coreNFW profile is defined as a radius-dependent suppression $f(r)$ of the NFW mass enclosed function $M_{\rm NFW}(r)$.  Thus to derive the stripped coreNFW profile, we simply applied the suppression $f(r)$ to a stripped NFW profile.  In reality, coreNFW halos should undergo more stripping than predicted by this approach, as cored halos are more susceptible to mass loss; $\sigma_{\rm los}^*$ will thus be slightly overestimated.  However, the differences in tidal stripping histories between cored and cusped halos only become apparent when the tidal radius is of order the core radius, corresponding to enclosed masses much less than 10\% of the total infall halo mass   \citep[e.g.,][]{2010MNRAS.406.1290P,2016MNRAS.461..710D,2018MNRAS.478.3879S}. Our treatment of tidal stripping is thus conservative.

A tidal stripping calibration also does not exist for our SIDM profile, and thus we adopt the following approach.  Firstly, we ignore the effects of non-gravitational interactions on halos after infall.  In other words, we do not take into account the possibility of increased core size (which grows with increased $t_{\rm age}$) and reduced density on account of self-interactions within the subhalo.  We also discount the possibility of expulsive or evaporative self-interactions between host halo and subhalo particles.  Both of these effects reduce the central density of the subhalo and accelerate satellite disruption \citep{2016MNRAS.461..710D}, so $\sigma_{\rm los}$ will be slightly overestimated.  For the range of cross-sections and the density profile we adopted, this is a good approximation for all but the most radial subhalo orbits.  Further, we consider cross sections small enough ($\sigma_{\rm SI}/m_\chi \lesssim 1 \hbox{ cm}^2/\hbox{g}$) that core-collapse via gravothermal catastrophe cannot occur within a Hubble time \citep[which typically requires a cross section $\sigma_{\rm SI} \gtrsim 10 \hbox{ cm}^2/\hbox{g}$;][]{2019arXiv190100499N}.  As we argue in Sec.~\ref{sec:sidm}, at large cross sections, most small subhalos would have to be far (but not too far) along the core-collapse process in order to reconcile the theoretical and observed VFs.

Outside $r_1$, we assume that the tidal stripping of the halo proceeds as the NFW case in \citet{2010MNRAS.406.1290P}, and then transitions within $r_1$ to a Burkert profile.  The density is continuous at $r_1$.  We test two different options for the evolution of the core formed by SIDM before infall.  First, we assume that the Burkert core radius $r_B$ is a fixed physical scale.  This is certainly too conservative---\citet{2016MNRAS.461..710D} found that the core size evolves more rapidly for SIDM cores than for CDM cusps for satellites of MW-sized hosts after infall.  Second, we allow $r_B$ and $r_1$ to evolve with the NFW scale radius $r_s$ so that the ratios $r_1/r_s$ and $r_B/r_s$ are fixed.  The two scenarios give nearly identical results unless the halos are stripped of substantially more than 90\% of their mass, at which point the satellites begin to be stripped of stars.

The effects of tidal stripping on $M(<r)$ and $\sigma_{\rm los}^*$ can be seen in Fig. \ref{fig:menc}.  There, as in the remainder of the paper, we show results for unstripped halos (m$_{\rm left}=1$), halos that have been 90\% stripped and where the stars may be starting to be stripped (m$_{\rm left}=0.1$), and halos that have been stripped of 99\% of their mass and may thus be considered close to dissolved in the case of cored halos \citep[m$_{\rm left}=0.01$;][]{2010PhRvD..82f3525S,2020MNRAS.491.4591E}.


\subsubsection{Evolution of the stellar distribution}
\label{sec:theory:stripping:stars}

As galaxies are embedded deep within their dark matter halos, they remain largely unaffected by tidal stripping until $\gtrsim$ 90\% of their dark matter halos has been stripped \citep{2008ApJ...673..226P,2018MNRAS.478.3879S}.  We model the change in the galaxy's structural parameters---and in particular, the half-light radius of the galaxy---as stripping proceeds as in \citet{2008ApJ...673..226P}.  Assuming dwarf galaxies follow a King profile, they provide fits for the evolution of its two key parameters, the King concentration $c_K$ and the core radius $r_c$, with mass loss.  From these parameters, we can compute the evolution of $R_{\rm eff}$. Including the evolution in $R_{\rm eff}$ causes a smaller decrement to $\sigma_{\rm los}^*$ than without.

We note that we do not evolve the stellar mass as a function of tidal stripping.  Although stellar mass loss slowly commences once $\gtrsim$ 90\% of the host dark matter halo is stripped, well over 99\% of the halo must be stripped before the stellar mass drops by a factor of 2 \citep{2008ApJ...672..904P, 2008ApJ...673..226P, 2020MNRAS.491.4591E}. We consider this level of stripping unlikely for the vast majority of the satellites we consider in this work, and is generally not captured in most high-resolution cosmological simulations \citep{2017MNRAS.468..885V, 2018MNRAS.474.3043V, 2018MNRAS.475.4066V, 2020arXiv201107077E}. In Sec.~\ref{sec:obs_uncert}, we consider how more extreme stripping of the observed nearby satellites might affect the construction of the completeness-corrected VF.


\begin{figure*}
    \centering
    \centerline{
    \includegraphics[width=0.5\textwidth]{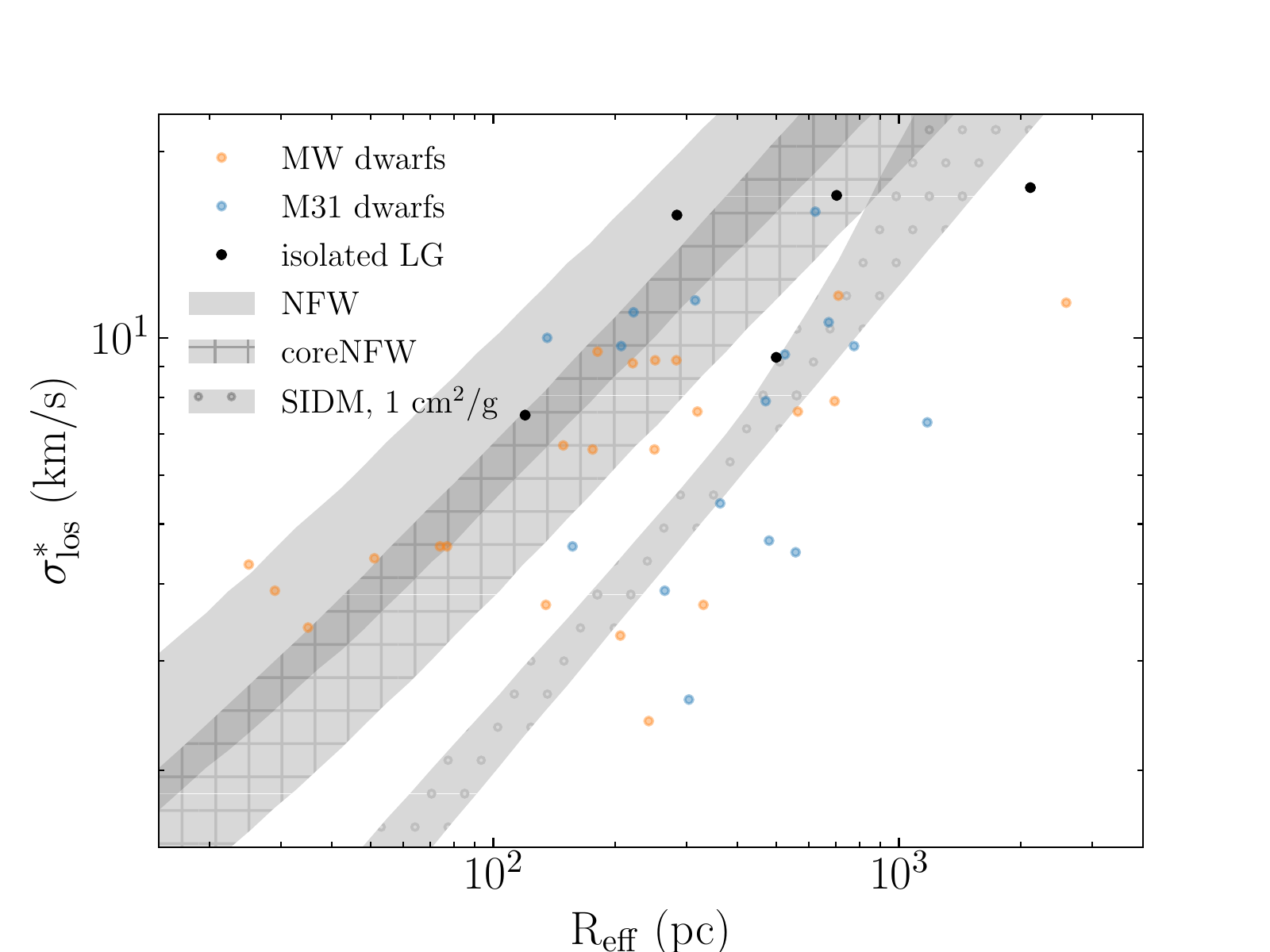}
    \includegraphics[width=0.5\textwidth]{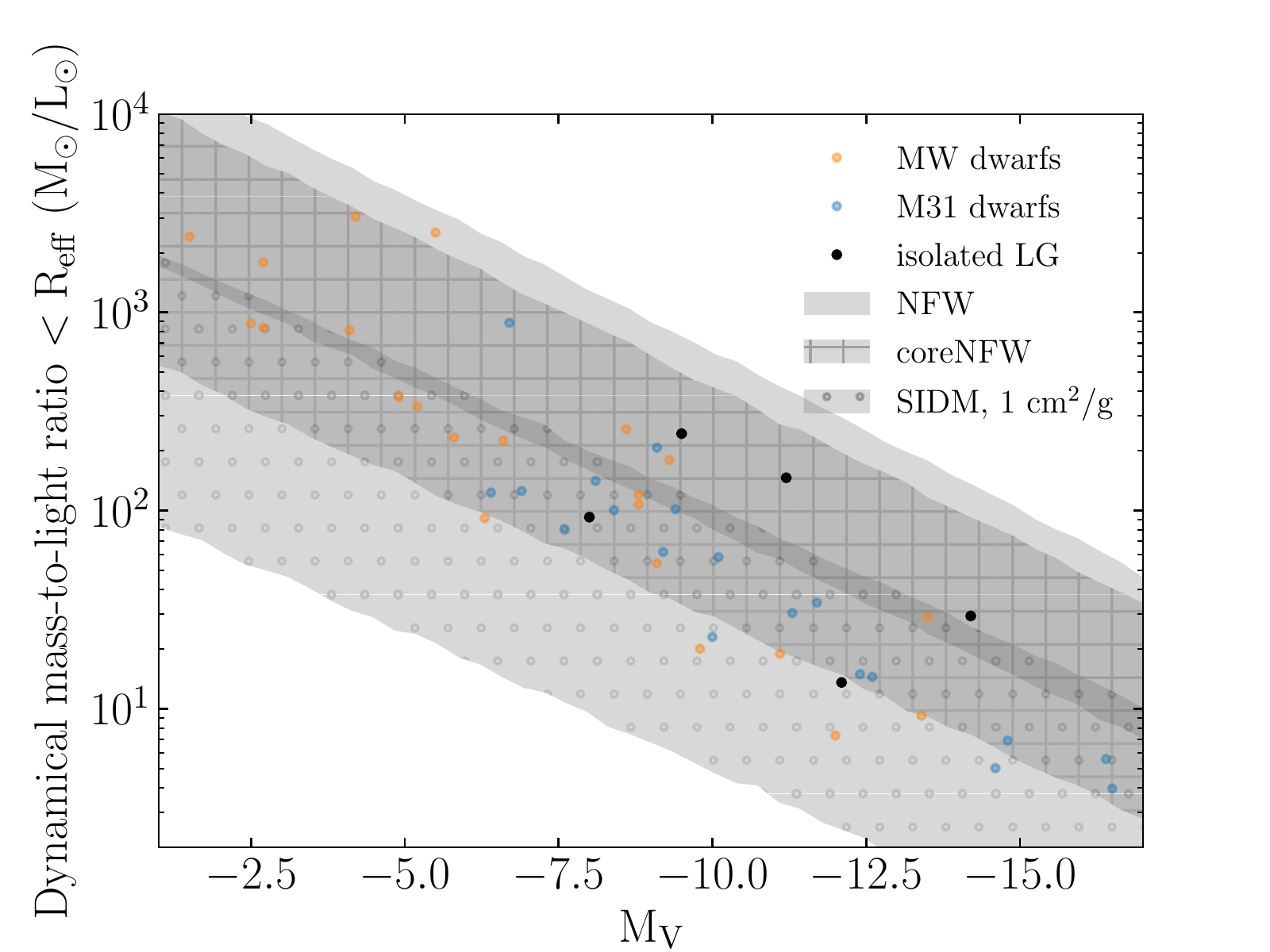}}
    \caption{Comparison of our model predictions (bands) against observations (points).  \emph{Left}:  The $\sigma^*_{\rm los}$-size relation for Local Group dwarfs.  \emph{Right}: The mass-to-light relation.  As in Fig. 2, the data were taken from \citet{2012AJ....144....4M}.  Each band corresponds to a different mass profile: NFW (no hatching), coreNFW (cross hatch), and SIDM with $\sigma_{\rm SI}/m_\chi = 1$ cm$^2$/g (dot hatch).  Bands include both a population of isolated dwarfs ($z_{\rm in}$=0) and unstripped dwarfs at the median infall redshift of the MW satellites, $z_{\rm in}$ = 1.  The width of the bands is dominated by the scatter in the SMHM, mass-concentration, and stellar size-mass relations we adopted, but also includes a small widening due to combining isolated and satellite populations.  Mass loss only has a significant effect on the properties plotted if $>$90\% of the halos are stripped and would lower both the stellar velocity dispersions (left) and the dynamical mass-to-light-ratios (right).  We calculate $M_V$ for each of the halos by calculating their stellar masses from the \citet{2013MNRAS.428.3121M} SMHM relation, dividing the result by a stellar mass-to-light ratio of 2 (appropriate for old stellar populations) to derive the luminosity, then converting it to a V-band magnitude.
    }
    \label{fig:checks}
\end{figure*}

\subsection{Comparisons against observed relations}

To demonstrate the ability of our model to reproduce the properties of observed Local Group dwarfs, we show our model predictions for observable properties in Fig. \ref{fig:checks}. In the left-hand panel, we plot the correlation between the 2D half-light radius vs. the velocity dispersion for both isolated and satellite Local Group dwarfs (points), along with our model predictions (bands).  In the right-hand panel, we show the dynamical mass-to-light ratio within the half-light radius vs. the V-band absolute magnitude.  For comparison against both isolated and satellite dwarfs, the bands include an isolated galaxy population (i.e. $z_{\rm in}=0$) and an unstripped dwarf population at the median infall redshift of the MW satellites, $z_{\rm in}=1$. The width of the bands is dominated by the scatter in the relations discussed throughout \S\ref{sec:theory} (in particular the SMHM, mass-concentration, and stellar mass-size relations), but also includes a small widening due to the superposition of isolated and satellite dwarf populations. For clarity, we do not show the evolutionary effects of tidal stripping.  As discussed above, unless $\gtrsim$ 90\% of each halo is stripped, the change in the predictions is minor \citep[e.g.][]{2018MNRAS.478.3879S,2020MNRAS.491.4591E}.

Of the three main density profiles we consider---NFW, coreNFW, and SIDM with $\sigma_{\rm SI}/m_\chi$ = 1 cm$^2$/g---the coreNFW model best matches the observational data.  The NFW model predicts velocity dispersions and dynamical mass-to-light ratios that are systematically higher at fixed $R_{\rm eff}$ or $M_{\rm V}$ than the coreNFW model.  In contrast, the SIDM model with $\sigma_{\rm SI}/m_\chi = 1$~cm$^2$/g is systematically low. A cross section of $\approx$ 0.1 cm$^2$/g leads to a better match with data.


\section{Results}\label{sec:results}

\begin{figure*}
    \centerline{
    \includegraphics[width=0.5\textwidth]{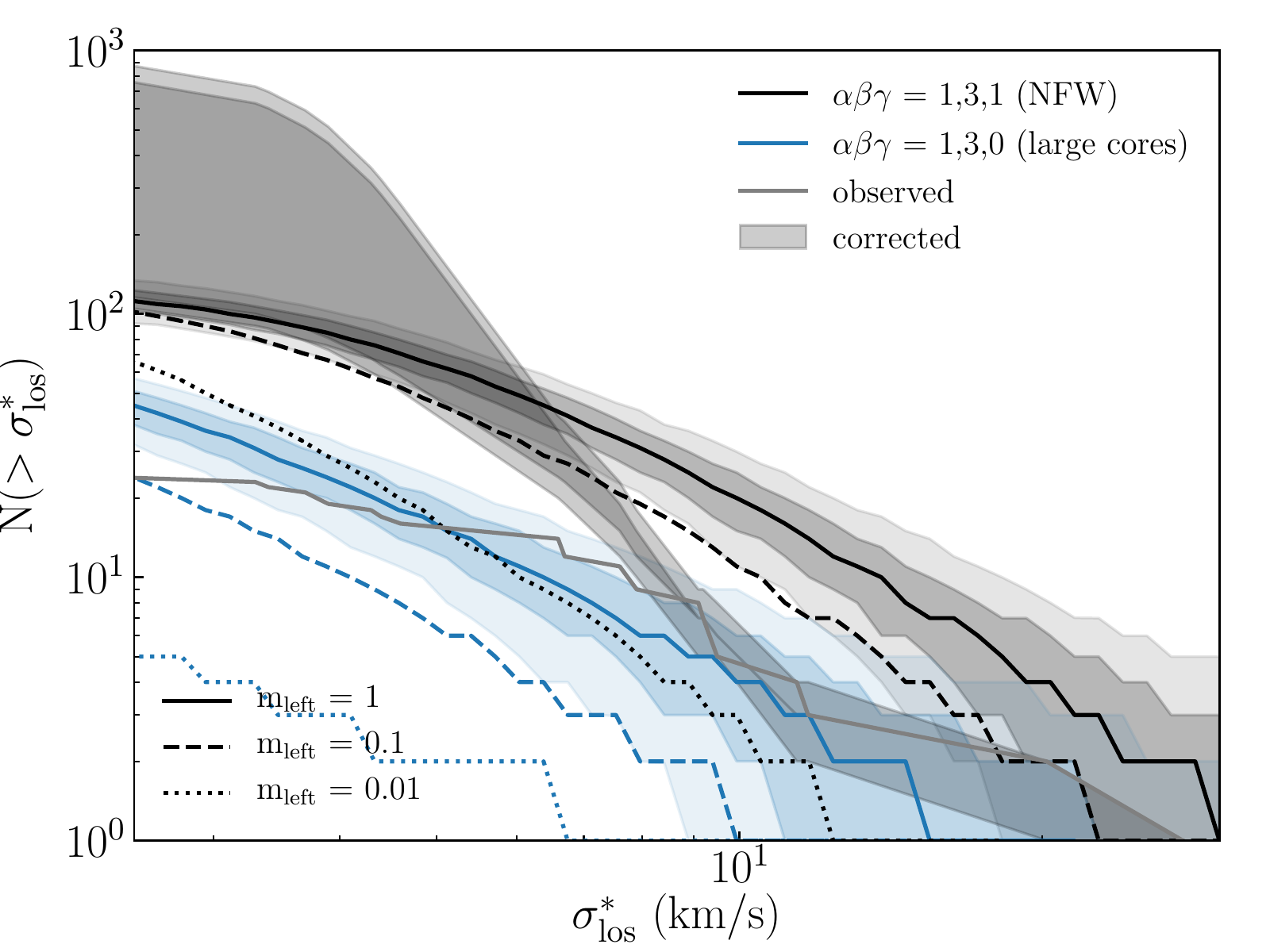}
    \includegraphics[width=0.5\textwidth]{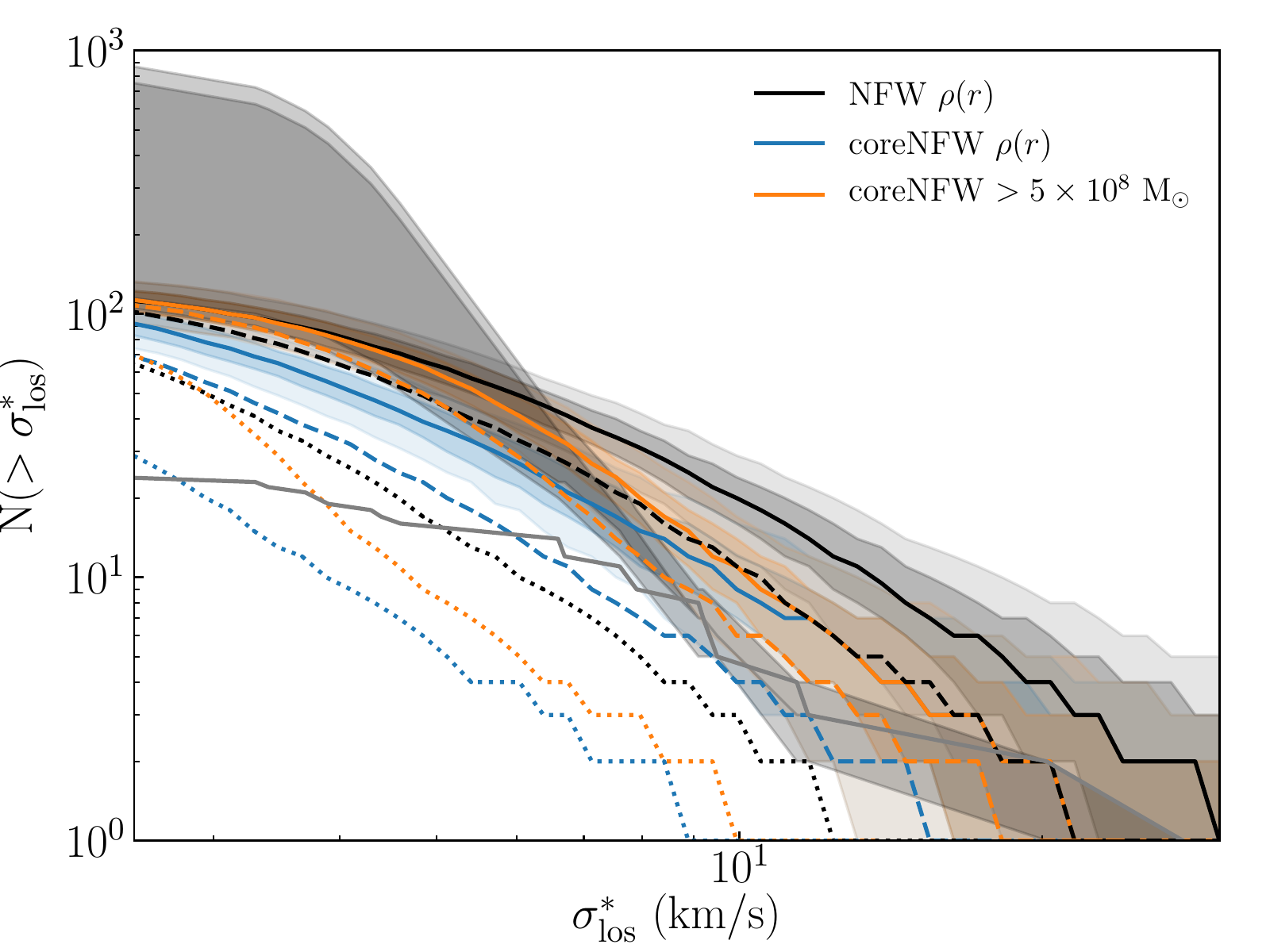}}
    \centerline{
    \includegraphics[width=0.5\textwidth]{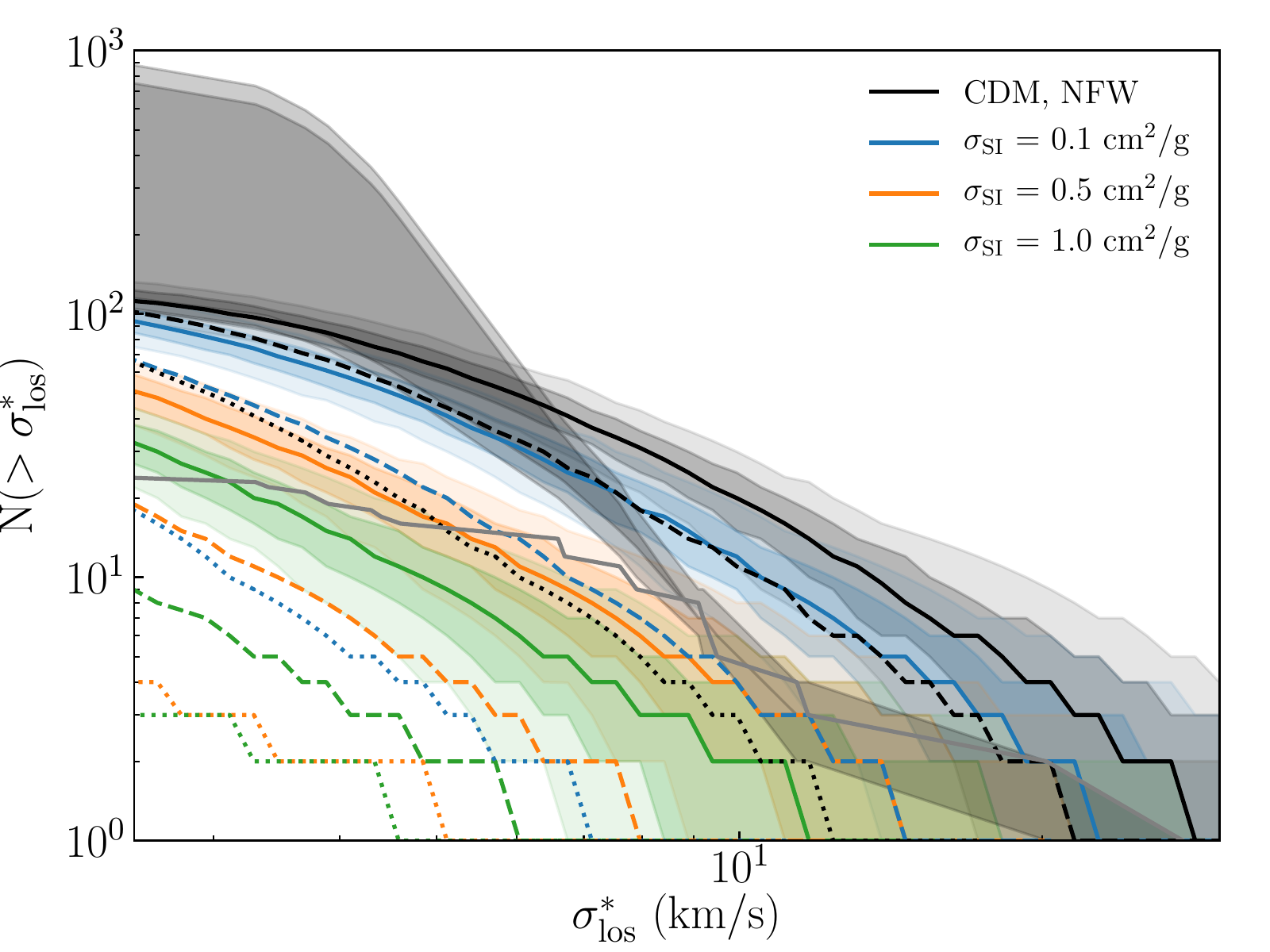}
    \includegraphics[width=0.5\textwidth]{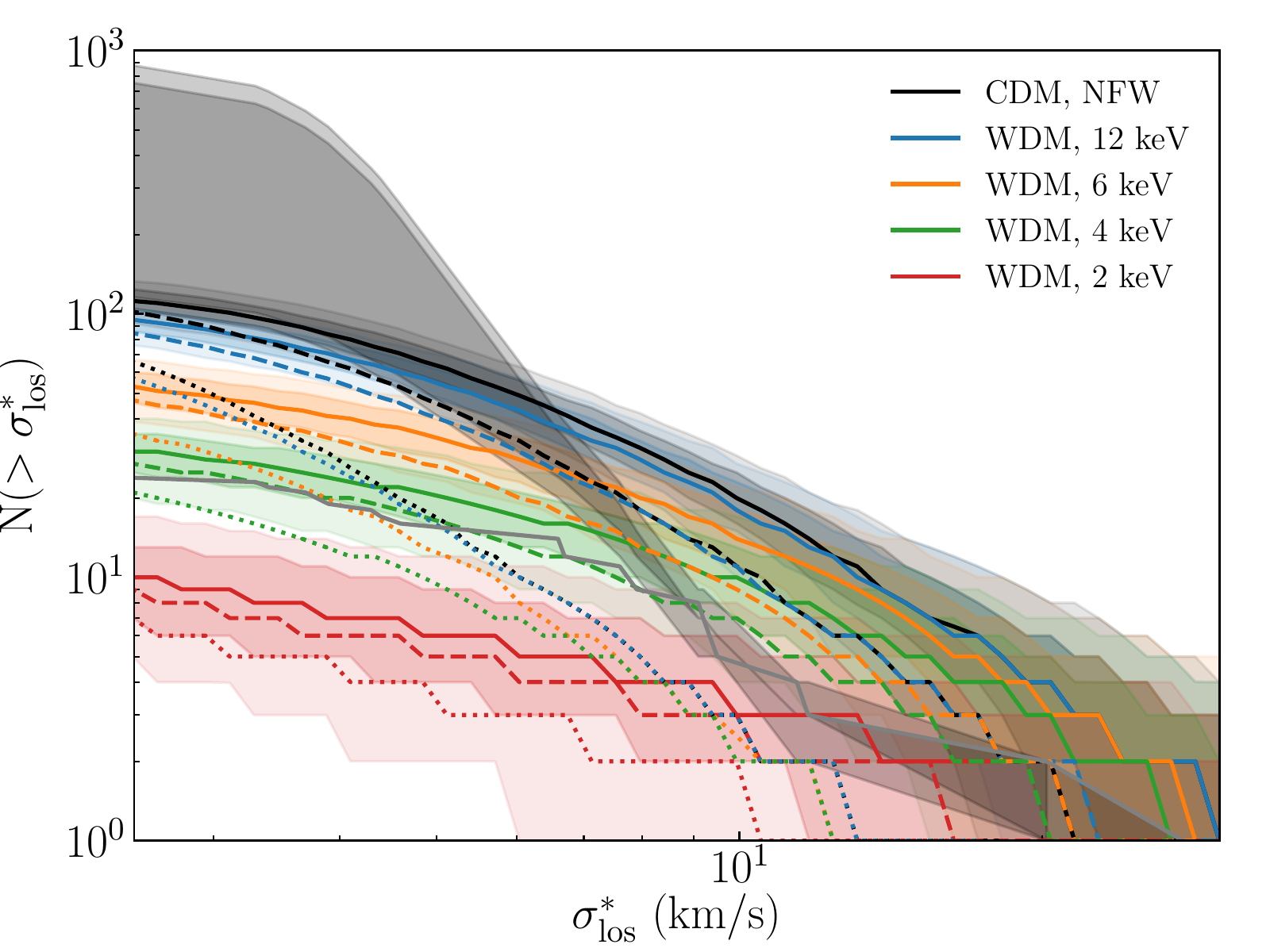}}
    \caption{Comparison of the theoretical (colored bands) and completeness-corrected (gray bands) velocity functions.  As before, the uncertainty on the completeness corrected VF show the 10-90\% percentile of satellite counts based on a Monte Carlo sampling of satellite $\sigma_{\rm los}$ given measurement uncertainties (assuming them to be Gaussian) as well as a 30\% scatter due to anisotropy.  The theoretical VFs are constructed using 1000 realizations of the MW satellite population, sampling the scatter in the stellar mass-size, stellar-mass--halo-mass, and mass-concentration relations, as well as in the halo-to-halo subhalo mass function.}
    \label{fig:vfxns}
\end{figure*}

Here we present comparisons of the completeness-corrected velocity functions shown in Fig. \ref{fig:vfxns-notheory} against the theoretical velocity functions for the MW satellites.  In Fig.~\ref{fig:vfxns}, we show velocity functions for each of the density profiles discussed in \S\ref{sec:density_profiles}.  In all panels, we show the theoretical velocity function predicted assuming no stripping (solid lines) and 10\% or 1\% of the dark matter mass left (dashed and dotted lines, respectively).  This range attempts to capture the uncertainty in the amount of tidal stripping experienced by MW satellites.  As discussed in Sec.~\ref{sec:obs_uncert}, while appropriate for the vast majority of satellites, the few with pericenters of order 10 kpc may be significantly more stripped \citep{2018MNRAS.478.3879S, 2020arXiv201107077E}.  Our goal in this section is to capture the typical population-level behavior of satellites, which is covered by the range of subhalo mass loss we adopt here.  The implications for the small-scale issues of the MW and for dark matter models are discussed below.


\subsection{Cores, cusps, and the Too Many Satellites problem} \label{sec:results:corecusp}

We start with comparing our observational completeness-corrected VFs against CDM predictions, both with and without baryons.  The top left of Fig. \ref{fig:vfxns} shows the theoretical VF predicted assuming all subhalos have an $\alpha\beta\gamma$ profile corresponding either to an NFW ($\alpha\beta\gamma$ = 1,3,1; black lines) or a cored ($\alpha\beta\gamma$ = 1,3,0; blue lines) profile.  The former is a baryon-free prediction for CDM halos, and the latter is an extreme case of how baryons might alter halos. The lines denote the median number of satellites in 1000 realizations of the MW satellite population.  The light and dark bands around these lines denote the 1 and 2$\sigma$ scatter due to the intrinsic scatter in the stellar mass-size, stellar-mass--halo-mass, and mass-concentration relations, as well as the halo-to-halo scatter in the subhalo mass function.

We first focus on the ultrafaint regime ($\lesssim$ 10 km/s), for which our completeness corrections are most uncertain on account of the unknown radial distribution of satellites. If subhalos are cuspy, the theoretical VF can explain the most conservative completeness-corrected VF (bottom edge of the shaded region), as long as the average satellite's halo is stripped by no more than 90\% of its dark matter.  This correction assumes that the spatial distribution of satellites follows the same NFW distribution as the smooth halo of the MW \citep{2016MNRAS.457.1208H,2021arXiv210301227G}, that visible satellites are recognized as galaxies despite significant stripping, and that there are no numerical issues with resolving satellites at the centers of their hosts \citep{2018PhRvL.121u1302K,Newton:2017xqg}.  

However, the correction assuming disk stripping (top edge of the shaded region) requires far more satellites than the theoretical predictions---a \emph{too many satellites} problem.  Both prosaic and exotic explanations for a too many satellites problem exist. In the former camp is the hypothesis that high-resolution cosmological simulations suffer from artificial disruption problems \citep[e.g.,][]{2018MNRAS.474.3043V,2018MNRAS.475.4066V,2020arXiv201107077E} which leads them to predict radial distributions of satellites that are not concentrated enough.  A more exotic interpretation is that the hydrodynamic simulations produce the correct radial distribution, but the matter power spectrum is boosted on small scales to reconcile the completeness-corrected VF with theory. 

At higher $\sigma_{\rm los}^*$, characteristic of classical satellites, we find a bend in the completeness-corrected VF, which is inconsistent with the theoretical VFs.  At intermediate $\sigma^*_{\rm los}$ $\sim$ 10 km/s, there is a significant gap between the completeness-corrected and theoretical VFs, which differ by as much as a factor of $\sim$2.  Unlike the issue of too many satellites at low velocities, at intermediate velocities we recover a missing satellites problem.  The discrepancy decreases---though does not completely disappear---if high-mass satellites undergo greater tidal stripping.  Simulations indicate that for more massive satellites, dynamical friction shortens their lifetimes and are destroyed more efficiently \citep{2020MNRAS.497.4459F}.

In contrast, assuming that all subhalos have an $\alpha\beta\gamma$ = 1,3,0 core profile results in a VF that struggles to reach the most conservative completeness-corrected VF, even if satellite halos are immune to tides.  This profile is thus strongly disfavored \citep[see also][who rule out big cores in ultrafaint satellites based on individual satellites' velocity dispersions]{2018MNRAS.481.5073E}.  As noted in \S\ref{sec:density_profiles}, this is because the cores are unrealistically large and diffuse for being baryonically created, producing unrealistically low $\sigma_{\rm los}^*$.

We find that density profiles arising from realistic models of baryonic feedback on dark-matter halos better capture the VF relative to the $\alpha\beta\gamma$ NFW or cored models, as shown in the top right panel of Fig. \ref{fig:vfxns}.  The blue line assumes all satellites have the largest baryonically-induced cores possible \citep[in the coreNFW model of][]{2016MNRAS.459.2573R}, i.e. assuming star formation continued until infall onto the MW at $z_{\rm in}$ = 1.  While this suppresses the theoretical VF relative to NFW, it is not as severe as the $\alpha\beta\gamma$ cored model discussed above, and in fact produces a closer match to the corrected VF than either at the higher $\sigma^*_{\rm los}$ end.  At the low $\sigma^*_{\rm los}$ end, however, only the most conservative completeness correction avoids the too many satellites regime, and only if the typical satellite halo is unaffected by tides.

It is unrealistic, however, to expect the smallest dwarfs to have cores.  Reionization suppresses star formation in ultrafaint dwarfs, and the absence of star formation-induced dark matter heating produces cusps \citep[though see][]{2021arXiv210102688O}.  We find that if we assume that all galaxies with $M_{200} \lesssim 5 \times 10^8$ M$_\odot$ have cusps, and those above have coreNFW profiles, we infer the VF denoted in orange on the second panel of Fig. \ref{fig:vfxns}.  The VF assuming different switch masses are shown in Appendix \ref{appdx:theory_uncert:mswitch}.  The switch mass $M_{200} \sim 5 \times 10^8$ M$_\odot$ corresponds to $\sim$7 km/s, or $M_* \sim 10^{3.5}$ M$_\odot$.  The switch from a core-dominated to a cusp-dominated regime  produces a better match to the shape and normalization of our most conservatively completeness-corrected VF, including the kink at $\sigma_{\rm los}^* \approx 10~\hbox{km/s}$, if the average satellite's dark matter subhalo is stripped by no more than 90\%. 

The threshold for the transition between cusps and cores are uncertain. Hydrodynamic simulations have not reached a consensus, with some placing it as high as $\sim$10$^{10}$ M$_\odot$ \citep{2014MNRAS.437..415D, 2016MNRAS.456.3542T, 2017MNRAS.472.2945R} or $\sim$10$^{9}$ M$_\odot$ \citep[i.e. $M_* \sim 10^5$ M$_\odot$;][]{governato2012, 2021arXiv210102688O}, while others find cores down to their resolution limit of 10$^8$ M$_\odot$ \citep{2016MNRAS.459.2573R}.  Observationally, it is either uncertain or impossible to ascertain for most MW satellites whether they possess a core or cusp given current observational facilities, but there is tentative evidence of a transition at $\sim$10 km/s, which corresponds to a threshold of $M_{200} \sim 10^9$ M$_\odot$ \citep{2018MNRAS.474.1398G, 2011ApJ...742...20W, 2012ApJ...746...89J, 2013MNRAS.429L..89A, 2001ApJ...563L.115K, 2013ApJ...763...91J, 2018MNRAS.481..860R, 2019MNRAS.484.1401R}.  Ultimately, there may be no single threshold mass \citep{2020MNRAS.497.2393L, 2019MNRAS.484.1401R}.  Any threshold that may exist only is distinguishable from the cores-all-the-way down case if tidal stripping is significant on a population level, or at very low velocity dispersions.

Observational uncertainties (\S\ref{sec:obs_uncert}) increase the significance of the gap at intermediate and high $\sigma_{\rm los}^*$.  In particular, the corrected VF assuming the older Boo II $\sigma_{\rm los}^*$ measurement can only be explained with cusps in dwarfs with high $\sigma_{\rm los}^*$.  Further, if dwarfs with signs of tidal disturbances are extremely tidally stripped, the corrected velocity function shows a sharp kink that is difficult to explain for any of the density profiles we have explored---even NFW fails to explain the number of corrected dwarfs with $\sigma_{\rm los}^*$ just below 10 km/s.

Overall, we find consistency with our previous work on the completeness-corrected \emph{luminosity} function \citep{2018PhRvL.121u1302K}, but the addition of velocity information adds significant distinguishing power for halo models.  As before, we find the number of completeness-corrected satellites can match the numbers predicted for CDM for the most conservative assumptions about the satellite distribution in the MW halo.  The VF further indicates that the smallest satellites ($\sigma_{\rm los}^* \lesssim 10~\hbox{km/s}$) are dense (either by cusps or small baryon-driven cores) and not be stripped by more than about 90\% of their halo mass, on average. Importantly, as in our previous work, we show that there may be a \emph{too many satellites} problem if the completeness corrections based on satellite distributions predicted by hydrodynamic simulations of tidal destruction by the MW's disk is accurate. For larger satellites with $\sigma_{\rm los}^* \gtrsim$ 10 km/s, baryon-produced cores are essential to matching the normalization and shape of the VF.

\subsection{SIDM}\label{sec:sidm}

The inability of large satellite cores to match the corrected VF can be used to place limits on dark matter models that predict significant coring.  We repeat the analysis in \S\S\ref{sec:ccVFs}-\ref{sec:theory} for one such model, SIDM.  With regards to the completeness corrections, hydrodynamic simulations indicate that at the current constraints for constant (velocity-independent) cross sections, $\sigma_{\rm SI}/m_\chi \lesssim$ 1 cm$^2$/g, the spatial distribution of satellites is similar to that in CDM \citep{2019MNRAS.490.2117R, 2019PhRvD.100f3007Z}.  As such, the corrections calculated in \S\ref{sec:ccVFs} for CDM are thus valid here.  We note that for larger cross sections (especially in the velocity-dependent case), larger SIDM-induced cores would increase satellites' susceptibility to tidal stripping and destruction near the host center, skewing the radial distribution of surviving satellites, and causing the completeness-corrected VFs to shift upwards.  This would exacerbate the too many satellites problem \citep[see, e.g.,][]{2016MNRAS.461..710D}.  With regards to the theoretical VF, we must input cored satellite density profiles appropriate for SIDM.  The density profile we adopted is discussed in \S\ref{sec:density_profiles}.

In the bottom left panel of Fig. \ref{fig:vfxns}, we show the VFs predicted for self-interaction cross sections ranging from $\sigma_{\rm SI}/m_\chi$ = 0.1 to 1 cm$^2$/g.  Self-interactions suppress the theoretical VF, even if the subhalo mass function is unchanged. While $\sigma_{\rm SI}/m_\chi \sim$ 0.5 cm$^2$/g induces sufficient suppression to resolve the mismatch between observed and theoretical VFs at intermediate $\sigma^*_{\rm los}$, it leads to a severe too many satellites problem at the low $\sigma^*_{\rm los}$ end.  Reproducing the most conservatively completeness-corrected VF at small $\sigma^*_{\rm los}$ requires a cross-section $\lesssim$ 0.1 cm$^2$/g. The difficulty that SIDM models with a constant $\sigma_{\rm SI}/m_\chi$ have in explaining the corrected VF is consistent with recent results showing that the diversity of MW stellar kinematics is difficult to reproduce with cross sections $\gtrsim$ 1 cm$^2$/g in the absence of core collapse  \citep[e.g.][]{2019PhRvX...9c1020R,2019PhRvD.100f3007Z,2021MNRAS.503..920C}.

While cuspier density profiles may be reestablished in SIDM subhalos due to tidal stripping-induced accelerated core-collapse \citep{2019arXiv190100499N,2019JCAP...12..010K}, consistency with the corrected VF requires a substantial fraction of subhalos to undergo core collapse  \citep{2019MNRAS.490..231K,2019PhRvD.100f3007Z,2020PhRvL.124n1102S}.  \citet{2021MNRAS.503..920C} suggests that a strongly velocity-dependent cross section is required to reproduce the pattern of central densities observed for the MW's classical dwarf galaxies.  It remains to be seen if the model is consistent with the ultrafaint dwarf population or with perturbations to their fiducial assumptions about the satellite subhalo initial conditions.  Small perturbations to the subhalo initial conditions can lead to large changes in the central density today on account of the runaway nature of core collapse \citep{2022MNRAS.513.4845Z}.  More broadly, it remains unknown whether velocity-dependent cross sections can induce just the right amount of accelerated core-collapse and reestablish cuspier density profiles among a significant fraction of ultrafaint-dwarf-hosting subhalos while keeping the central densities of classical-dwarf-hosting subhalos low.


\begin{figure*}
    \centering
    \centerline{
    \includegraphics[width=0.5\textwidth]{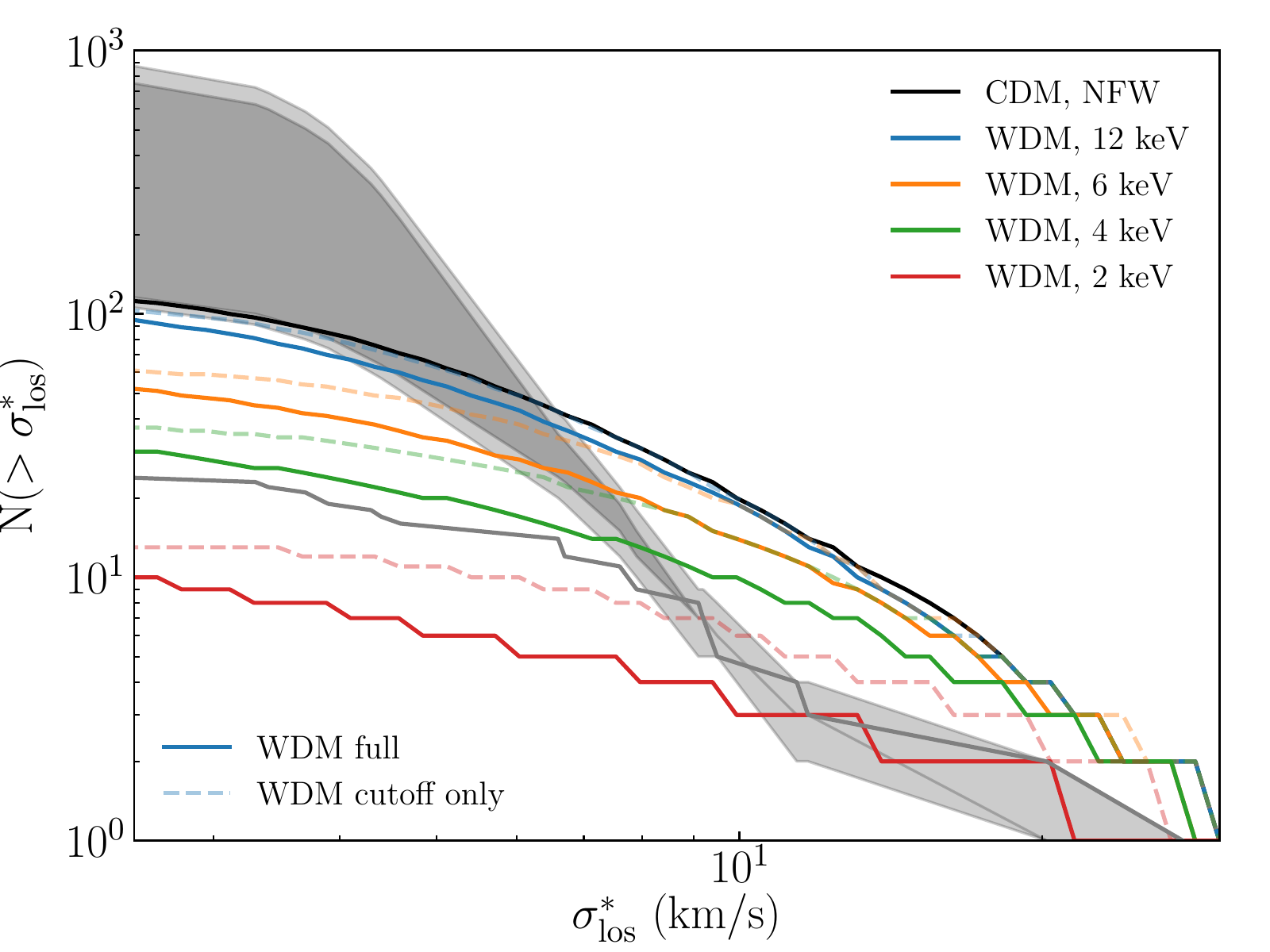}
    \includegraphics[width=0.5\textwidth]{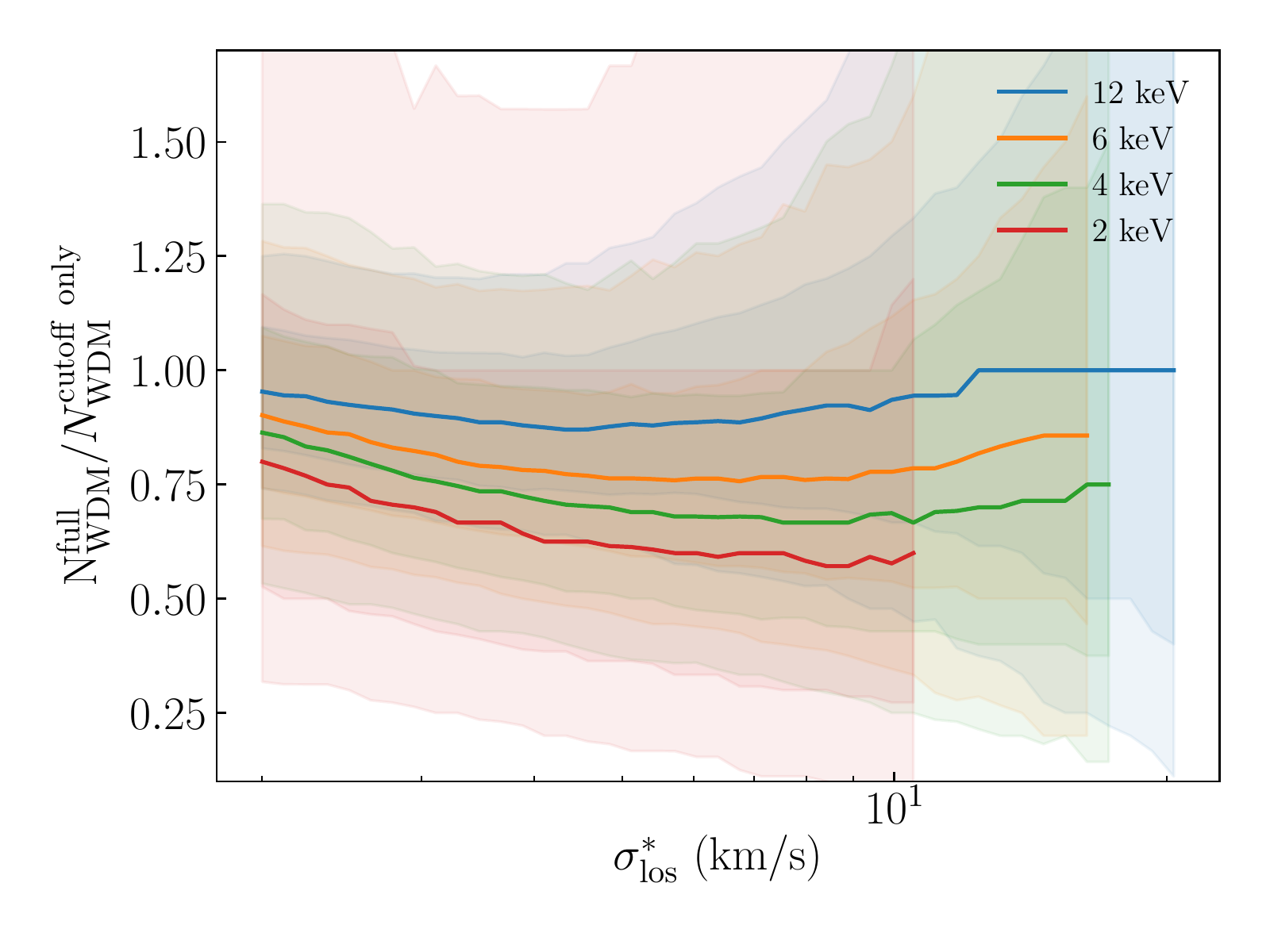}}
    \caption{The impact of including WDM concentrations on the velocity function.  (\emph{Left}) The solid lines show the full WDM model, which includes both the cutoff in the WDM mass function, as well as lower WDM concentrations.  The light dashed lines only include the cutoff in the WDM mass function.  Including WDM concentrations further lowers the VF, producing a tighter constraint on WDM than with the cutoff alone. (\emph{Right}) The suppression in number counts with just the cutoff alone vs with lower WDM concentrations.  The median suppression and the 1- and 2-$\sigma$ contours from our 1000 realizations of a WDM MW satellite population are shown in the lines and colored bands, respectively.  A 4 keV thermal relic results in an additional 15-30\% additional reduction in the VF by including lower concentrations.  The VF of a 6 keV thermal relic is additionally reduced by 10-20\%.}
    \label{fig:vfxns-wdm}
\end{figure*}

\subsection{WDM}\label{sec:wdm}

The satellite VF is unique probe of warm dark matter (WDM) models due to its sensitivity to two key features of WDM:  the suppression of power and internal densities at low mass scales. The VF's sensitivity to both the mass function and density profile makes it a stronger probe of WDM than the satellite luminosity function alone \citep{2014MNRAS.442.2487K, 2017arXiv170804247N, 2017MNRAS.464.4520B, 2018MNRAS.473.2060J, 2018PhRvL.121u1302K, 2019ApJ...873...34N, 2020arXiv201108865N, 2020arXiv201013802E, 2021PhRvL.126i1101N, 2021arXiv210107810N}. Suppression of small-scale power is a classic feature of WDM models, and as halos form later in a WDM cosmology \citep[e.g.][]{2013MNRAS.428.1774B}, they have lower central densities than in CDM.  We model the former by adopting the conversion from CDM to WDM mass functions given in \citet{2020ApJ...897..147L}.  The structure of WDM halos have been found to only slightly deviate from NFW profiles, but have a different mass-concentration relation due to their later formation \citep{2008ApJ...673..203C, 2014MNRAS.439..300L, 2016MNRAS.460.1214L}.  We adopt the relation given in \citet{2012MNRAS.424..684S}.  Again, hydrodynamical simulations indicate that the radial distribution of WDM satellites is similar to that in CDM \citep{2014MNRAS.439..300L}, so the completeness corrections calculated in \S\ref{sec:ccVFs} are valid for WDM as well.

The resulting VF is shown in the bottom right panel of Fig. \ref{fig:vfxns} for WDM thermal relics with masses from 2 to 12 keV.  As expected, there is significant suppression at the low $\sigma^*_{\rm los}$ end.  Reproducing the most conservatively completeness-corrected number of satellites requires a thermal relic with a mass of at least 6 keV.  This is consistent with recent luminosity-function-based constraints with the Milky Way satellites recently discovered in the Dark Energy Survey \citep{2020arXiv201013802E,2021PhRvL.126i1101N,2021arXiv210107810N,2020arXiv201108865N}.  We reiterate that the bands on the WDM VFs only reflect the intrinsic scatter in the scaling relations we adopted, and do not include the uncertainty in the MW mass---including the uncertainty would significantly increase the size of the bands.  The constraint on the WDM mass depends sensitively on the MW mass \citep{2014MNRAS.442.2487K}, for which we have adopted a single fiducial value of 10$^{12}$ M$_\odot$.  On the higher $\sigma^*_{\rm los}$ end, there is little suppression in the VF.  All models over-predict the number of intermediate-mass satellites, potentially indicating the need to include baryonic cores in massive WDM satellites, which have been shown in WDM hydrodynamic simulations to improve the match to MW satellite kinematics \citep{2017MNRAS.468.4285L, 2017MNRAS.468.2836L,2019MNRAS.483.4086B}.

Constraints on WDM from the VF are stronger than one may obtain from the luminosity function alone, as shown in Fig.~\ref{fig:vfxns-wdm}.  In the left panel of this figure, we show VFs for WDM models with our full treatment of the power spectrum suppression (solid lines), and the case in which the halo mass function is suppressed but the halo concentrations are assumed to follow the CDM relation (dashed lines).  The later formation time of WDM halos relative to CDM suppresses the VF significantly.  Constraints on WDM with the luminosity function are based on the suppression of the halo mass function, but not of the concentration.  Recently, several groups showed that constraints on WDM from combining substructure lensing (sensitive to both halo abundance and density) with MW satellite populations (tracing halo abundance) are tighter than considering either observable alone \citep{2020arXiv201013802E,2021arXiv210107810N}.  Our work similarly indicates that stronger constraints may be possible with the VF than with satellite abundances.  We calculated the suppression in the VF with both the WDM abundance and concentration suppression relative to to the VF with just the abundance suppression alone for each of our 1000 realizations of the WDM MW VF.  In the right panel of Fig.~\ref{fig:vfxns-wdm}, we show the median (line) and 1- and 2-$\sigma$ contours of the additional suppression (bands) induced by lower concentrations for each of the thermal relic masses we considered. We find that the WDM concentration additionally suppresses the velocity function by a percentage that depends both on the WDM thermal relic mass and on $\sigma^*_{\rm los}$.  At the 4 keV thermal relic mass limit derived from jointly constraining the MW satellite luminosity function and strongly lensed substructures, we find the suppression to be 10\%-30\%. There is significant scatter at the high $\sigma_{\rm los}^*$ and for low-mass thermal relics due to Poisson noise.

\subsection{Summary}

We have shown that it is possible for the MW satellite VF to match the simplest CDM theoretical predictions if completeness corrections are taken into account.  However, our fiducial theoretical model suggests the following is required in order to match the observed VF: 
\begin{itemize}[leftmargin=0.5cm]
\item The ultrafaint galaxies ($\sigma^*_{\rm los} \lesssim 10$ km/s) have dense, cuspy halos, in line with other work on their central densities \citep{2018MNRAS.481.5073E,2019PhRvD.100f3007Z}.  
\item Tidal stripping must not have reduced the mass of the average ultrafaint galaxy's dark matter halo by much more than $\sim$90\%.
\item The classical dwarf galaxies' VF ($\sigma^*_{\rm los} \gtrsim 10$ km/s) is better matched by halos with baryon-driven cores than to cuspy halos, although we still overpredict the VF unless subhalos are stripped of $\sim 90$\% of their dark matter. 
\item Dwarf satellites must largely survive close to the center of the MW, or else such large completeness corrections apply that we have a too many satellites problem \citep[see also][]{2018PhRvL.121u1302K,2019MNRAS.tmp.1496K}.
\end{itemize}
We note, however, that the overall shape of the VF is difficult to reproduce with our fiducial model (notably near $\sigma^*_{\rm los} \sim 10$ km/s).  It becomes even more challenging if the radial distribution of satellites is less centrally concentrated than our most conservative assumption, or if the observational VF resembles the two alternative completeness correction scenarios discussed in \S\ref{sec:obs_uncert}.

More broadly, the VF adds distinguishing power for alternatives to the CDM paradigm for dark matter over satellite counts alone.  Satellite luminosity functions are governed largely by the subhalo mass function, while the VF encodes additional information about the density profile of halos.  For the simplest SIDM models, the subhalo mass function and radial distribution is similar to those in CDM, but the VF is significantly affected by the effects of dark-matter scatter on the central density.  For WDM, both the suppression in the subhalo mass function and the lower central density of subhalos leave their mark on the VF.  In future work, we will present quantitative constraints on SIDM and WDM models with the VF framework we describe in this work.


\section{Discussion}\label{sec:discussion}

Although there is broad agreement between the completeness-corrected VF and predictions from CDM theory, we have shown that there is some difficulty in reproducing the shape of the VF, particularly near $\sigma_{\rm los}^* \sim$ 10 km/s, and that we struggle to produce theoretical VFs with more satellites than the most conservatively completeness-corrected VF.  In this section, we test if well-motivated alternatives to our fiducial theoretical assumptions in \S\ref{sec:theory} can better match the observed VF.  Furthermore, we are interested in exploring whether alternative theoretical choices can produce a VF that exceeds the most completeness-corrected VF at small $\sigma_{\rm los}^*$, and/or accommodate the other observational VF completeness corrections outlined in \S\ref{sec:obs_uncert}.  More broadly, we demonstrate the robustness of our main conclusions:  that CDM with a set of standard assumptions of galaxy-halo scaling relations provides a good explanation for the observational completeness-corrected VF of MW satellite galaxies.

\subsection{Reproducing the shape of the VF}

At the heart of the mismatch between the corrected and theoretical VFs is a discrepancy in the \emph{shape} of the respective VFs:  the most conservative correction can be matched at low and high $\sigma_{\rm los}^*$ ends, but not at $\sim$10 km/s.  One could imagine attempting to reduce the normalization of the theoretical VF to produce match at 10 km/s, but this would result in a too many satellites problem at low $\sigma_{\rm los}^*$ and underestimate the VF at high $\sigma_{\rm los}^*$.

Many variations to our fiducial theoretical model mainly affect the \emph{normalization} of the VF.  As we show in Appendix \ref{appdx:theory_uncert}, these include alternative mass-concentration relations, stellar-mass--galaxy-size relations, and subhalo infall times.  Adopting these alternative relations involve replacing the fiducial power-law scaling relation with another.  Changes in the \emph{shape}, in addition to the normalization, of the theoretical VF are models can be achieved by adopting scaling relations that depart from a power-law form.  We have already encountered one such example in \S\ref{sec:results:corecusp}: a switch between cusps and cores, which introduces a steeper VF slope at intermediate $\sigma_{\rm los}^*$.  Here, we discuss an additional means to introduce scale-breaking relations that could provide a better match to the corrected VF:  the stellar-mass--halo-mass relation. 

\begin{figure}
    \includegraphics[width=0.5\textwidth]{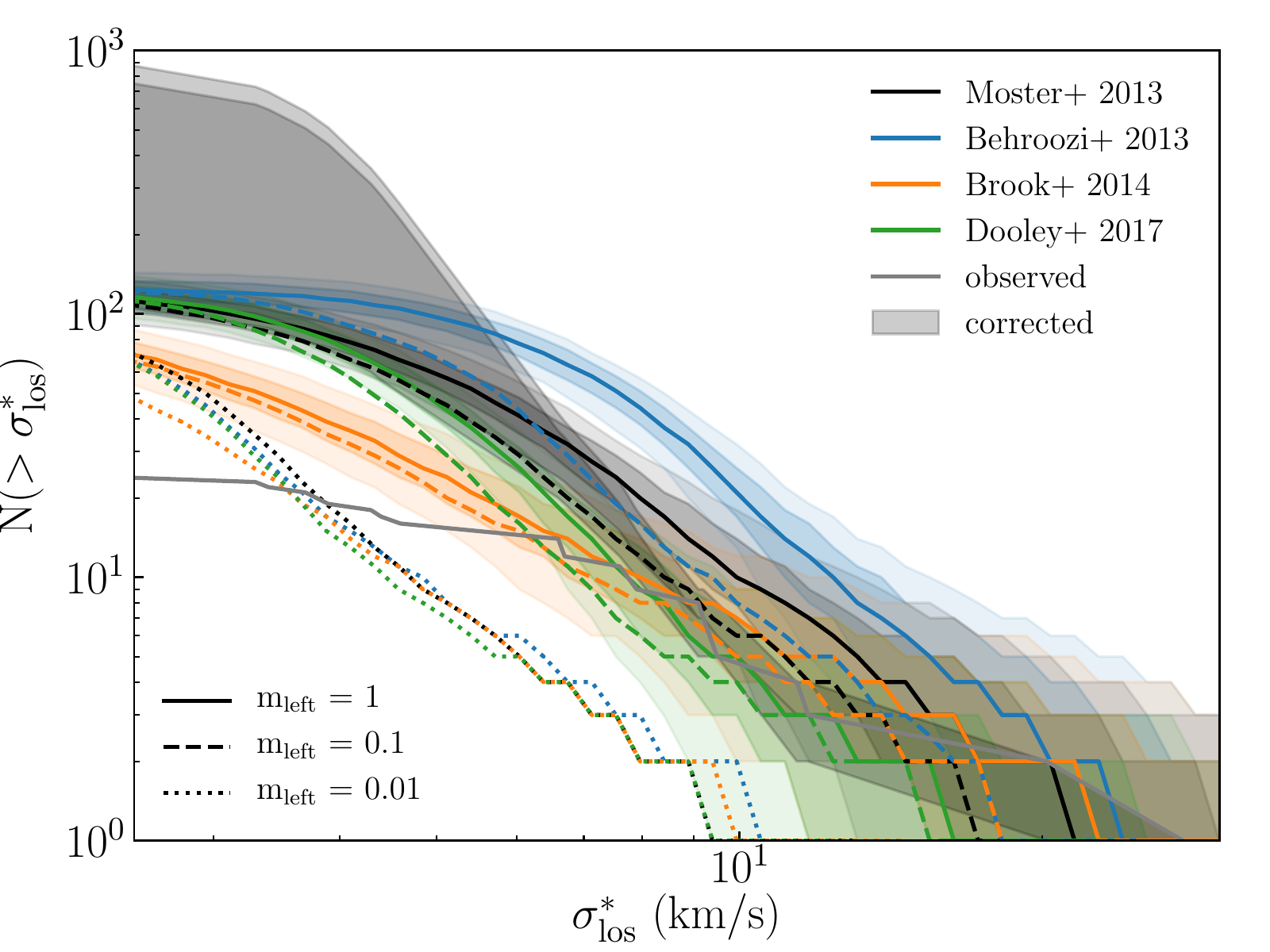}
    \caption{Same as Fig. \ref{fig:vfxns}, but exploring different SMHMs, which can change the shape/slope of the theoretical VF.  Fiducial choices are plotted in black.  Satellites with masses $M_{200} < 5 \times 10^8$ M$_\odot$ were assumed to have coreNFW density profiles.  A SMHM with a power-law break, such as the one by \citet{2017MNRAS.471.4894D} that predicts that the SMHM flattens at low halo masses, produces a remarkably good match to the most conservatively completeness-corrected VF.}
    \label{fig:vfxn_matching_shape}
\end{figure}

The SMHM relation is required to determine the stellar mass of each halo, which enters our calculation for $\sigma_{\rm los}^*$ in two places:  in calculating the stellar contribution to $M(<R_{\rm eff})$, as well as in computing $R_{\rm eff}$ from the stellar mass-size relation  (see \S\ref{sec:theory}).  However, the SMHM relation is poorly constrained at the scale of ultrafaint dwarfs, at which the relations in the literature span as much as two orders of magnitude \citep[e.g.,][]{2017MNRAS.471.4894D, 2018ARA&A..56..435W, 2021arXiv210105822M}.  Indeed, most of the relations we discuss here have been calibrated at higher mass scales and are extrapolations below $M_{200} \sim 10^9$ M$_\odot$.  Further, each individual relation exhibits significant scatter, growing from about 0.3 dex at scales above $\sim$10$^{10}$ M$_\odot$ to over a dex at scales of $\sim$10$^8$ M$_\odot$ \citep{2017MNRAS.464.3108G,2019MNRAS.483.1314B,2021arXiv210504560G,2021arXiv210105822M}.  

In Fig. \ref{fig:vfxn_matching_shape}, we show the theoretical VF derived using different SMHM relations.  We assume that density profiles below an infall halo mass of $5\times 10^8 M_\odot$ are cusped, and those above are described by the coreNFW profile.   Our fiducial SMHM relation is from \citet{2013MNRAS.428.3121M} (black), who derived a 0.15 dex intrinsic scatter when fitting their relation, which only included halos down to 10$^9$ M$_\odot$. In comparison, we show the VF assuming a SMHM with a severe power-law break from \citet[][see also \cite{2015MNRAS.448.2941S}]{2017MNRAS.472.1060D} in green. This relation takes the form of a double power law that \emph{flattens} at masses below $2 \times 10^9$ M$_\odot$.  This flattening leads to a slight suppression in the VF at high $\sigma_{\rm los}^*$, followed by significant steepening in the VF at $\sim$10 km/s, producing a remarkable match to the shape of the most conservatively completeness-corrected VF.  This break in the SMHM is expected at some scale---galaxies and star clusters cannot be arbitrarily small in stellar mass.  Our work potentially hints at a scale.

We note that the converse---a steepening of the SMHM relation, such as in \cite{2021arXiv210105822M}, which predicts a double power law with a break at halo scales of around $10^{10}$ M$_\odot$---causes the opposite effect as expected, suppressing the theoretical VF at low masses.  While \cite{2021arXiv210105822M}'s relation remains consistent with the corrected VF, a more severe steepening would underpredict it, producing a too many satellites problem even for the most conservative completeness corrections.

We also show two other relations that bracket the range of SMHMs in the literature.  Alternative SMHM relations that predict a larger $M_*$ for a given halo mass, such as \citet{2013ApJ...770...57B}, shown in blue, would predict a larger $R_{\rm eff}$, $M(<R_{\rm eff})$, and thus a larger $\sigma_{\rm los}^*$.  Adopting the \citet{2013ApJ...770...57B} relation overshoots the corrected VF, increasing the mismatch at $\sim$10 km/s and introducing a severe missing satellites problem at that scale.  Consistency with the corrected VF requires dwarfs at the $\sim$10 km/s scale to be $>$90\% stripped.  Given that most dwarfs do not seem to be undergoing stellar stripping, this does not seem likely \citep{2008ApJ...673..226P, 2015MNRAS.449L..46E}.  Conversely, SMHM relations that predict smaller $M_*$ for a given halo mass, such as \citet{2013MNRAS.428.3121M}, shown in orange, result in smaller $R_{\rm eff}$, $M(<R_{\rm eff})$, and thus smaller $\sigma_{\rm los}^*$.  The \citet{2014ApJ...784L..14B} relation cannot produce even the most conservatively completeness-corrected number of satellites at the low velocity end.  

None of the SMHM relations help resolve the too many satellites problem should the corrections accounting for disk stripping be correct.

Last but not least, we note that not all breaks in the scaling relations may clearly manifest in the VF.  For example, the switch between cores in classical and cusps in ultrafaint dwarfs in \S\ref{sec:results:corecusp}---a break in the central density scaling relation---introduces a feature that mirrors the sharp change in slope of the observed VF at $\sigma_{\rm LOS}^* \sim$ 10 km/s.  However, the scatter in the mass-concentration and the stellar mass-size relations erases this feature in the VF (see Appendix \ref{appdx:theory_uncert:scatter}). Thus, a break in a scaling relation is a necessary condition to introduce features in the VF, but the strength of the break is mediated by scatter.

\subsection{Matching VFs under large completeness corrections}

In \S\ref{sec:ccVFs}, we showed that larger completeness corrections (associated with greater tidal stripping) predict as many as several 100s to a few 1000s of satellites, particularly those with low $\sigma_{\rm los}^*$.  If such corrections are accurate, this poses a problem---none of the variations on our fiducial model discussed thus far have had the ability to produce such numbers.  Here, we discuss two additional variations to our fiducial model that predict greater numbers of satellites: later reionization and a larger Milky Way mass.

\begin{figure*}
    \centerline{
    \includegraphics[width=0.5\textwidth]{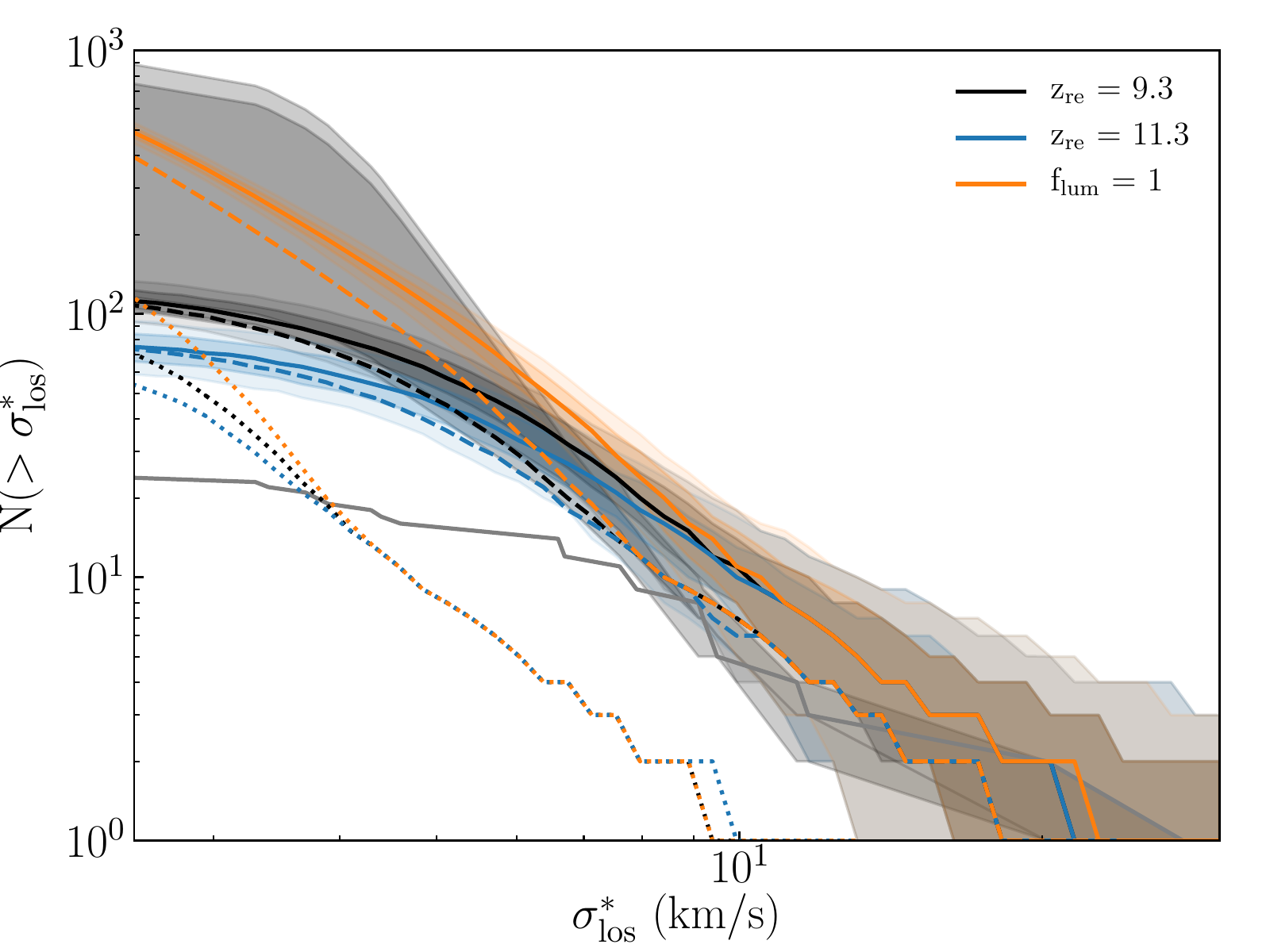}
    \includegraphics[width=0.5\textwidth]{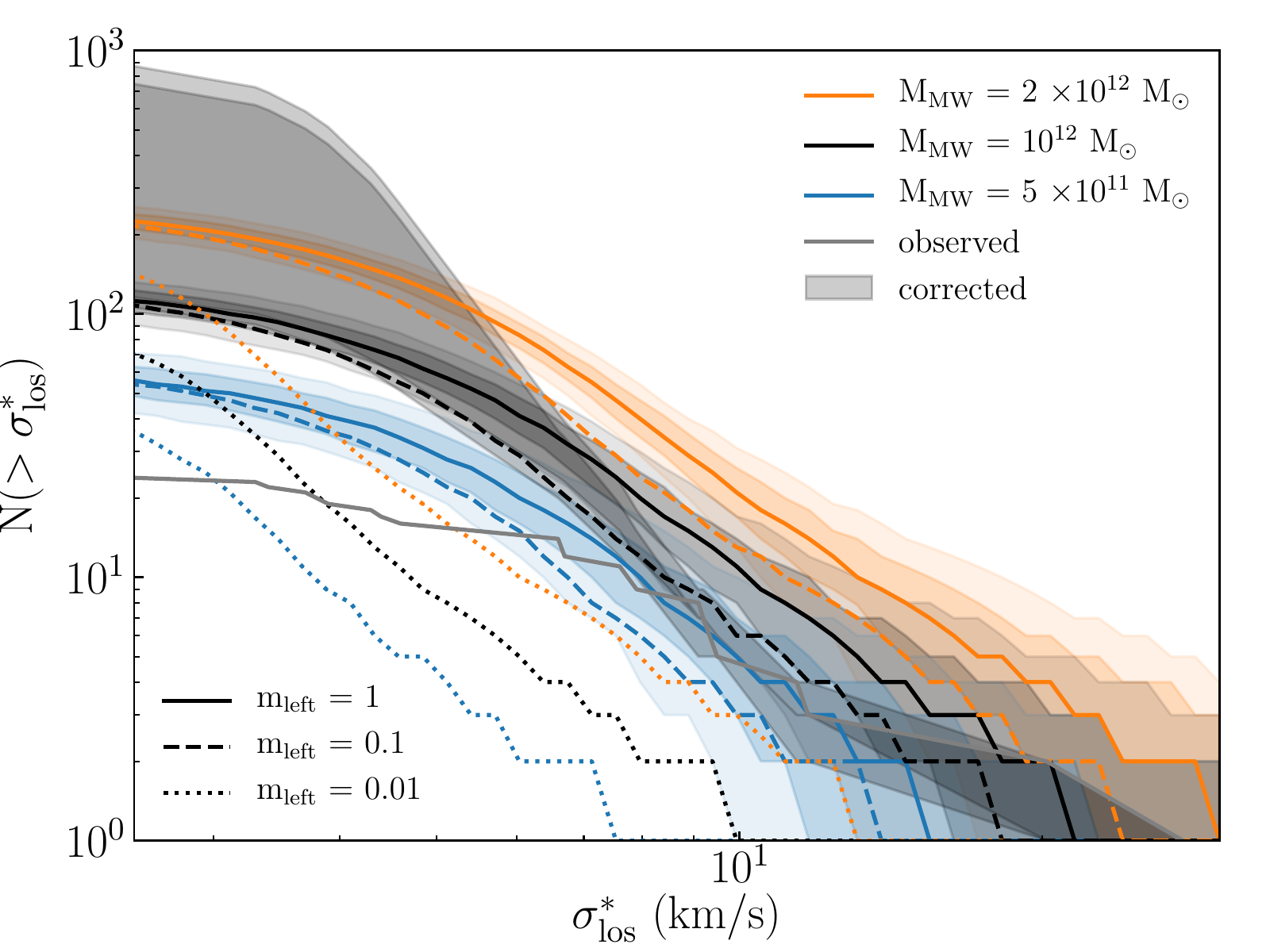}}
    \caption{Same as Fig. \ref{fig:vfxns}, but exploring theoretical uncertainties that could help produce a better match to corrected velocity functions with large completeness corrections.  Fiducial choices are plotted in black.  Satellites with masses $M_{200} < 5 \times 10^8$ M$_\odot$ were assumed to have coreNFW density profiles.  (\emph{Left}) Theoretical VFs with varying reionization redshifts, which affects the fraction of subhalos that do not host a galaxy at a given mass.  Later reionization can produce more low-mass galaxies.  (\emph{Right}) Theoretical VFs with the MW mass spanning its factor of two uncertainty.  While a larger MW mass can produce more galaxies, it overpredicts the corrected VF at higher velocity scales.  Conversely, a better fit at $\sigma_{\rm los}^* \gtrsim$ 10 km/s can be achieved with a lower MW mass, but results in a too many satellites problem at the low $\sigma_{\rm los}^*$ end.}
    \label{fig:vfxn_matching_high_ccVFs}
\end{figure*}

\subsubsection{Reionization redshift}

Another break in scaling relations that we adopted in our fiducial theoretical model was introduced in \S\ref{sec:theory}:  a cutoff in the galaxy mass function due to reionization.  In the right-hand panel of Fig. \ref{fig:vfxn_matching_high_ccVFs}, we show in black the theoretical VF derived using our fiducial reionization model.  Our fiducial reionization redshift, $z_{\rm re}$ = 9.3, is consistent with Planck 2015 results \citep{2016A&A...596A.108P}, but is earlier than recent measurements suggest \citep[e.g.][]{Aghanim:2018eyx}.  Later reionization would allow more halos to grow and reach the mass threshold to form galaxies, increasing the number of observable MW satellites.  However, as noted in \S\ref{sec:theory}, we adopted a generous ``go" criterion to populate halos with galaxies, using the dual threshold model of \citet{2017MNRAS.471.4894D}.  This model assumes a galaxy-formation threshold of $v_{\rm max} > 9.5$ km/s for halos before reionization, which is between the molecular and atomic cooling scales, and a post-reionization threshold $v_{\rm max} > 23.5$ km/s, which is lower than the typical $v_{\rm max} \gtrsim 30 - 50$ km/s after reionization \citep{2000ApJ...542..535G,2008MNRAS.390..920O}.  Thus our generous criterion allows our results to be conservatively applied to later reionization redshifts. 

In the left-hand panel of Fig. \ref{fig:vfxn_matching_high_ccVFs}, we show the effect of removing this scale-breaking cutoff.  The fiducial $z_{\rm re} = 9.3$ model is shown in black.  In orange is the maximal number of galaxies that could be observed, if all halos hosted galaxies.  Removing the break induced by reionization produces a nearly power-law form for the VF and predicts hundreds of low-mass galaxies, significantly reducing the severity of the too many satellites problem at the low $\sigma_{\rm los}^*$ end, if large completeness corrections are accurate.  Note however, that even if all halos hosted galaxies, we cannot reproduce the largest completeness-corrected number of satellites.  In contrast, if we assume reionization takes place even earlier than our fiducial model, at $z_{\rm re}$ = 11.3, it produces the extra suppression shown in blue.  Modifying the reionization redshift only affects the VF at low velocities.  This is in line with the expectation that reionization introduces a sharp transition at subhalo masses between 10$^{8-9}$ M$_\odot$ (corresponding to $\sigma_{\rm los}^* \sim$ few km/s) above which subhalos host galaxies, and below which they do not \citep{2018MNRAS.473.2060J, 2018PhRvL.121u1302K, 2021PhRvL.126i1101N}.

\subsubsection{The Milky Way's mass}

The number of MW galaxies also depends on the mass of the MW, as the number of satellites above a fixed mass scales linearly with the host halo's mass.  There is a factor of $\sim$2 uncertainty in the mass of the MW, ranging from 0.5 - 2 $\times 10^{12}$ M$_\odot$ \citep[e.g.][]{2020SCPMA..6309801W}.  Our fiducial mass M$_{\rm MW} = 10^{12}$ M$_\odot$ lies in the middle of this range.  We show how the VF shifts assuming the range of uncertainty in the MW's mass in the right-hand panel of Fig. \ref{fig:vfxn_matching_high_ccVFs}.  As expected, a more massive MW would help partially alleviate the too many satellites problem posed by large completeness corrections.  However, unless the satellites were, on average, at least 90\% stripped and/or a significant number of ultrafaint dwarfs were cored, a more massive MW would reintroduce the missing satellites problem and exacerbate the mismatch at $\sim$10 km/s even with cored massive satellites.  A less massive MW, on the other hand, would avoid the tension at $\sim$10 km/s \citep{2013MNRAS.428.1696V} but introduce a severe too many satellites problem, rule out cored ultrafaint galaxies, and require minimal stripping, although this could be alleviated by later reionization. The fiducial 10$^{12}$ M$_\odot$ MW best matches the observed VF.

\subsection{Summary}

Within the context of CDM, we can model the feature in the completeness-corrected VF at $\sigma_{\rm los}^* \sim$ 10 km/s and produce VFs that match or exceed the most conservative completeness correction at low $\sigma_{\rm los}^*$.  We can achieve the feature if there is a sharp break in the SMHM relation as proposed by \citet{2017MNRAS.472.1060D}. However, sharp breaks in the mean scaling relations are softened in the predicted VF on account of intrinsic scatter in all scaling relations relating to halo and galaxy properties.  We can exceed the most conservative completeness-corrected VF at low $\sigma_{\rm los}^*$ the mass of the MW is significantly higher than our fiducial value of $M_h = 10^{12} M_\odot$, at the cost of exacerbating the missing satellites problem at the classical dwarf scale.   The high-MW-mass solution is additionally disfavored by recent analysis of the influence of the Large Magellanic Cloud on dynamical mass measurements of the MW \citep{2020MNRAS.498.5574E,2021MNRAS.501.5964D}.  Alternatively, we can increase the low-$\sigma_{\rm los}^*$ end of the VF if reionization is late \citep[in line with recent measurements;][]{2016A&A...596A.108P}, but even if all subhalos are occupied by luminous baryons, we are unable to reach the upper edge of the completeness-corrected VF band.  Reaching the upper end of this band may require an increase to the inflationary power spectrum on small scales \cite[see, e.g.,][]{2022MNRAS.512.3163G}.  In short, the observed VF can be modeled in terms of reasonable CDM + galaxy formation models, unless the true VF is close to the top end of our observed completeness-corrected VF uncertainty band.


\section{Conclusions}
\label{sec:resolution}

The satellite galaxy velocity function encodes information about the physics that shapes the subhalo mass function, subhalo structure, and how galaxies inhabit these halos.  In this work, we present observationally based completeness-corrected MW satellite velocity functions, and compare them to theoretical predictions.  Our theoretical framework, based on scaling relations and their scatter, allows us to efficiently explore the many uncertainties in galaxy formation and dark matter physics, highlight how these affect our interpretation of the MW satellite velocity function, and lay out a path to sharper constraints. 

Our key result is that we find good consistency between our fiducial theoretical CDM+baryon velocity function and the most conservative completeness-corrected velocity function, presented in Fig.~\ref{fig:vfxns}.  The shape of the velocity function suggests a switch from baryon-driven cores for satellites $\gtrsim 5 \times 10^8$ M$_\odot$ to cusps for lower mass satellites, in line with simulations and observational work.  The shape of the velocity function around 10 km/s may also be indicative of a break in the SMHM relation \citep{2017MNRAS.472.1060D} and relatively recent reionization redshift \citep{2016A&A...596A.108P}.  Our most conservatively completeness-corrected velocity function assumes that galaxies are not yet transformed into tidal debris even if their halos are highly stripped, as recent simulations suggest \citep{Newton:2017xqg,2018MNRAS.478.3879S,2020arXiv201107077E,2021arXiv210504560G}.  On the other hand, if galaxies are destroyed by the MW's disk as in cosmological hydrodynamic simulations \citep{2010ApJ...709.1138D,2013ApJ...765...22B,2017MNRAS.471.1709G,2020MNRAS.492.5780R,2020MNRAS.491.1471S}, then the observed abundance of dwarf galaxies close to the center of the MW's halo pushes us toward a \emph{too many satellites} problem \citep{2018PhRvL.121u1302K}. Further, assuming such stripping produces a completeness-corrected velocity function that vastly overshoots even our most optimistic predictions for satellites with low velocity dispersion.  Neither a large mass for the MW nor allowing every subhalo to host a luminous galaxy can yield a sufficiently high theoretical VF at small velocities.  This may indicate that high-resolution cosmological simulations suffer from artificial disruption problems \citep[e.g.,][]{2018MNRAS.474.3043V,2018MNRAS.475.4066V,2020arXiv201107077E} or that the matter power spectrum is boosted on small scales.

We use our framework to show how the observed VF constrains non-CDM models of dark matter.  Given that our fiducial CDM + baryon model is consistent with conservative corrections, dark matter models that either reduce power on small scales or produce large cores in halos are tightly constrained.  For SIDM models below the core-collapse regime, we find that velocity-independent elastic scattering cross sections are restricted to $\sigma_{\rm SI}/m_\chi \lesssim 0.5$~cm$^2$/g.  For larger cross sections, the larger cores lead to a suppressed velocity function for a fixed subhalo mass function.  In future work, it will be important to explore the velocity-dependent cross section regime in which core collapse is possible \citep{2019JCAP...12..010K,2019MNRAS.490..231K,2020PhRvL.124n1102S,2021MNRAS.503..920C}.  The challenge for those models is to produce dense cores in ultrafaint dwarf galaxies while preserving the relatively low density of the classical satellites.  We find that a thermal relic WDM candidate must have a mass $\gtrsim$ 6 keV to reproduce the low-$\sigma^{*}_{\rm los}$ end of the velocity function assuming a MW mass of 10$^{12}$ M$_\odot$, comparable to constraints from other recent work \citep{2017PhRvD..96b3522I, 2020arXiv201013802E,2021PhRvL.126i1101N,2021arXiv210107810N}.  Constraints on WDM with the velocity function are tighter than with satellite counts alone because the velocity function is suppressed both by the reduction in the subhalo mass function and by the lower central density arising from later halo formation times.  Going forward, we strongly endorse combining luminosity and velocity measurements of MW satellites in WDM analyses.

There are several paths forward to a more robust measurement of the MW's satellite velocity function, and a more precise interpretation.  First, we showed in Sec.~\ref{sec:obs_uncert} that shifts in velocity dispersion measurements of individual nearby ultrafaint dwarf galaxies can significantly alter the shape of the velocity function.  Because non-power-law features in the velocity function imply significant breaks in empirical galaxy-halo scaling relations, reproducing the shape of the velocity function is key to its theoretical interpretation.  Thus, better control over statistical and (more critically) systematic uncertainties is essential. Second, assumptions about the tidal state of the known nearby dwarf galaxies affect how we assign velocity dispersions to the unobserved, more distant, similar-luminosity analogs of the nearby ultrafaint galaxies  that dominate the completeness correction \citep{2011AJ....142..128W,2013ApJ...770...16K,2017MNRAS.467..573C,2018MNRAS.476.3816F,2019ApJ...885...53M, 2020ApJ...902..106M}.  More accurate models of tidal evolution for individual satellites will increase the accuracy of completeness corrections.  Finally, the discovery of new and distant satellites with the Vera C. Rubin Observatory \citep{2014ApJ...795L..13H,2018PhRvL.121u1302K,2019ApJ...873...34N,2019arXiv190201055D}, and the measurement of their velocity dispersions \citep{simon2019baas} will enable a more direct measurement of the velocity function.  Because of observational incompleteness, the highly uncertain spatial distribution of satellite galaxies in the halos of Milky Way-like systems is our single-greatest source of uncertainty in the observed Milky Way velocity function \citep{2018PhRvL.121u1302K,2020ApJ...902..124C}.  Fundamentally, this uncertainty has its origins in the fate of highly stripped halos in even very high-resolution simulations \citep{2020arXiv201107077E}. The spatial distribution can be measured observationally, and the observational incompleteness mitigated, by the sensitivity of the Vera C. Rubin Observatory to distant satellites, as demonstrated by discoveries with the Hyper Suprime-Cam Subaru Strategic Program  \citep{2016ApJ...832...21H,2018PASJ...70S..18H,2019PASJ...71...94H}.  Furthermore, next-generation facilities like Rubin and spectroscopic facilities on the extremely large telescope class of observatories, will enable us to extend our analyses to systems beyond the MW.

The power of satellite velocity functions to illuminate dark-matter and galaxy-evolution physics harkens back to the early days of the missing satellites problem \citep{1999ApJ...522...82K,1999ApJ...524L..19M}.  When sharper measurements of satellite velocity dispersions are paired with a full statistical analysis and a better understanding of galaxy formation at small scales, velocity functions will be powerful tests of dark matter physics at the edge of galaxy formation for the MW and beyond.


\section*{Acknowledgements}
We thank Chris Kochanek, David Weinberg, and Justin Read for helpful comments.  SYK acknowledges support from the Ohio State University Graduate School through the Presidential Fellowship. AHGP acknowledges support from National Science Foundation (NSF) Grant No. AST-1615838 \& AST-1813628. This material is based on work additionally supported by NASA under award number 80NSSC18K1014. 

\section*{Data Availability}

Data available upon request.


\bibliographystyle{mnras}
\bibliography{mw_vf}



\appendix

\section{Non-monotonicity of density profile evolution with \citet{2010MNRAS.406.1290P}}
\label{appendix:notes_on_P10}

\begin{figure}
    \includegraphics[width=0.5\textwidth]{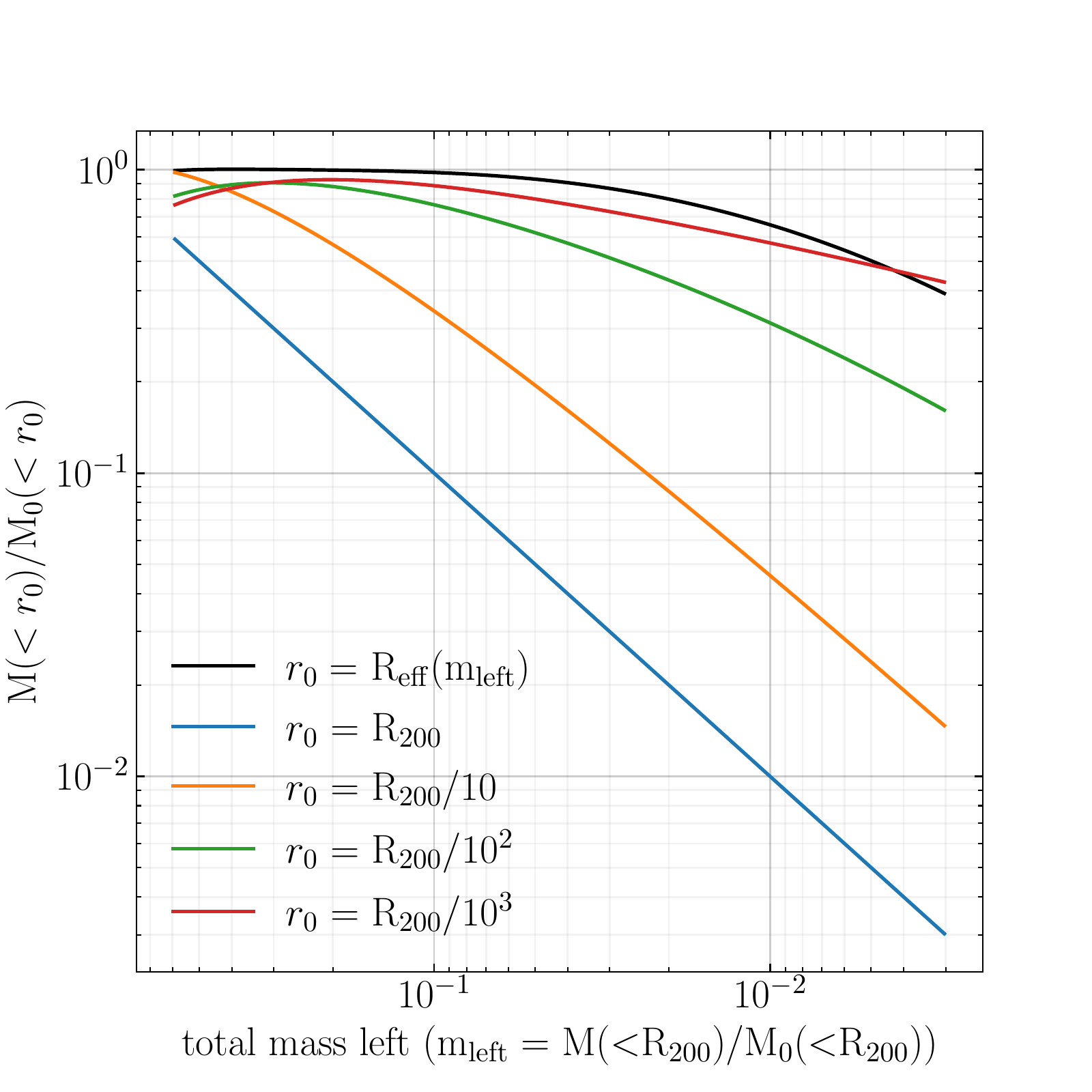}
    \caption{Enclosed mass within a given radius (fraction of the original virial radius) as a function of mass lost from a subhalo with total mass M$_{200} = 10^8$ M$_\odot$.  Note that for small radii $\le R_{200}/100$, the enclosed mass initially increases.  However, this does not occur within the half-light radius R$_{\rm eff}$ (black), which evolves as tidal stripping progresses, and which is the enclosed mass relevant to our work.  The total mass left (m$_{\rm left}$) starts at $\sim$60\% because \citet{2010MNRAS.406.1290P}'s density profiles are valid for m$_{\rm left}^{100} \le$ 0.9 (e.g. the mass left as defined in the $\Delta \sim 100$ system), which corresponds to m$_{\rm left}^{200} \sim 0.6$.}
    \label{fig:menc_vs_time}
\end{figure}

\begin{figure}
    \includegraphics[width=0.5\textwidth,clip,trim=0.6cm 0.6cm 1.85cm 1.47cm]{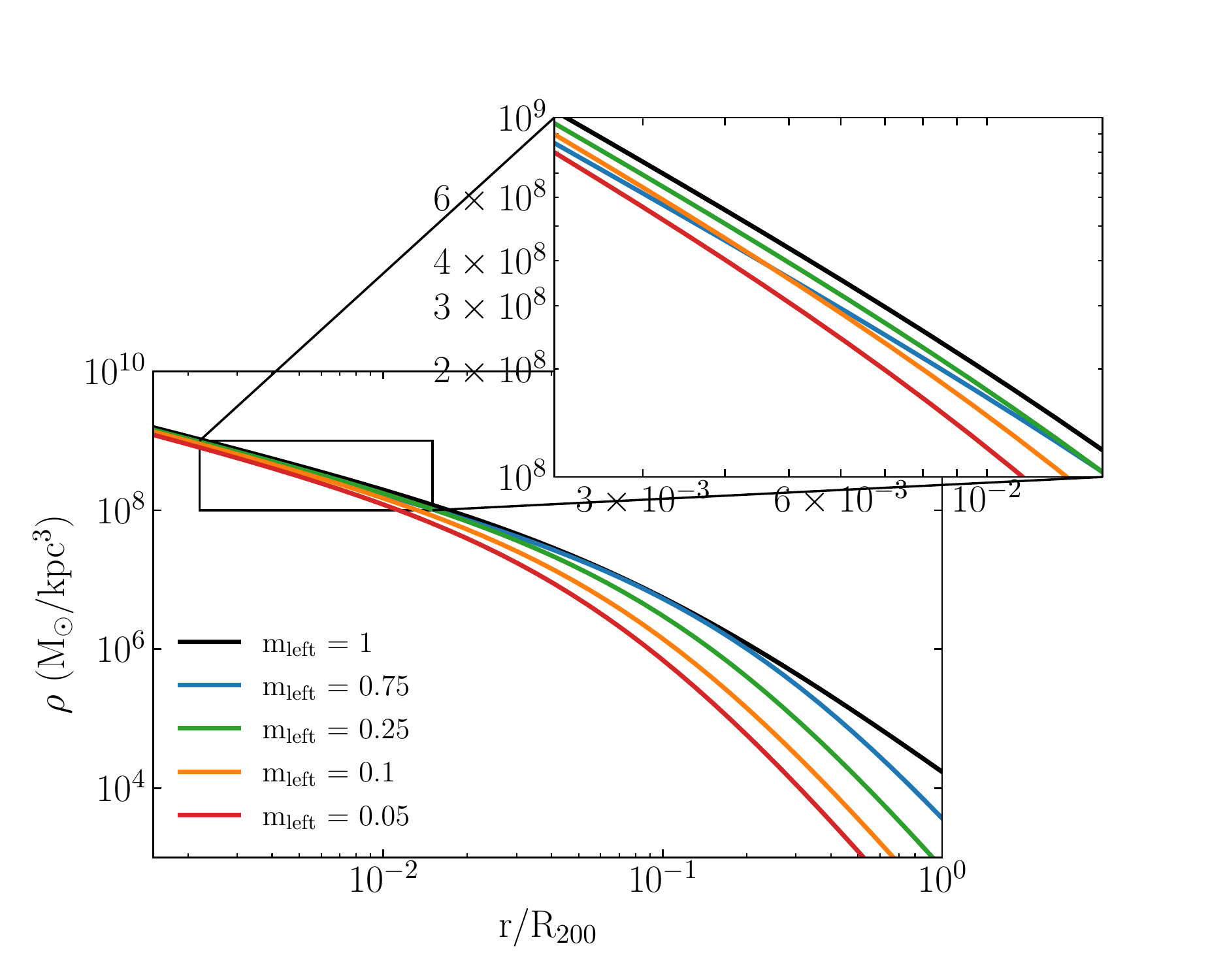}
    \caption{Density profile for the same 10$^8$ M$_\odot$ subhalo as in Fig. \ref{fig:menc_vs_time} before tidal stripping (black) and with 75\% (blue), 25\% (green), 10\% (orange), or 5\% (red) of its mass left.  While at large radii, the density drops with mass loss, the inset shows that at very small radii $\le 10^{-2}$ R$_{100}$, the density increases before decreasing again.  This causes the non-monotonic change in the enclosed mass at small radii shown in Fig. \ref{fig:menc_vs_time}.}
    \label{fig:rho_vs_time}
\end{figure}

To model the effect of tidal stripping on the density profiles of the MW satellites, we used the prescription in Appendix A3 of \citet{2010MNRAS.406.1290P}.  Simply put, stripping causes density profiles to fall as $r^{-5}$ in their outskirts, and changes in their structural parameters are dictated by the total amount of mass lost from the subhalo.

While tidal stripping removes mass from all radii, we found that this prescription does not predict a monotonic decrease in the amount of mass within very small fixed radii ($r <$ R$_{200}/100$), as shown in Fig. \ref{fig:menc_vs_time}.  The mass within small radii \emph{increases} as tidal stripping progresses (and thus the total bound mass $m_{\rm left}$ decreases) before falling monotonically as expected.  We find that this is due to the behavior of the stripped density profile at small radii, as shown in Fig. \ref{fig:rho_vs_time}.  While at radii $>$ R$_{200}/100$, densities lower monotonically with increasing mass loss, this is not true at smaller radii.  Densities at radii $<$ R$_{200}/100$ \emph{increase} when the subhalo's mass drops from 90\% to about 20-50\% of its original mass.  This causes the increase in the enclosed mass exhibited in Fig. \ref{fig:menc_vs_time}.  We checked that at the larger mass loss rates explored by \citet{2010MNRAS.406.1290P}, we reproduce their density profiles (shown in their Fig. 5).  

We note that our application of this prescription pushes beyond the range of radii and m$_{\rm left}$ considered in \citet{2010MNRAS.406.1290P}.  The radii we considered are at the half-light radii of dwarf galaxies, which span R$_{200}/1000 < R_{\rm eff} <$ R$_{200}/100$, and included intermediate mass loss (m$_{\rm left} > 0.1$), while the authors generally considered larger amounts of stripping, m$_{\rm left} < 0.2$.  The prescription may for this reason not be best calibrated at the scales of interest to this work.

However, we note that in our work, instead of adopting a \emph{fixed} radius, we adopted an \emph{evolving} half-light radius that changes as tidal stripping progresses as given by \citet{2008ApJ...673..226P} and discussed in \S\ref{sec:theory:stripping:stars}.  The mass enclosed by this evolving radius does not exhibit this initial increase in mass (i.e. black line in Fig. \ref{fig:menc_vs_time}), and thus this behavior does not appear in our work.


\section{Additional sources of uncertainty in the theoretical VF}
\label{appdx:theory_uncert}

In \S\ref{sec:discussion}, we explored variations on theoretical assumptions that produced changes in the shape of the VF at $\sim$10 km/s to better match the completeness-corrected VF.  We also explored variations that predicted greater numbers of satellites with low $\sigma_{\rm los}^*$ that is required if large completeness-corrections are accurate.  Here, we explore other variations to our fiducial theoretical model that---despite having significant scatter or uncertainties that impact other analyses and/or are interesting in their own right---do not affect the theoretical VF, and thus our main conclusions.

\begin{figure}
    \centering
    \includegraphics[width=0.5\textwidth]{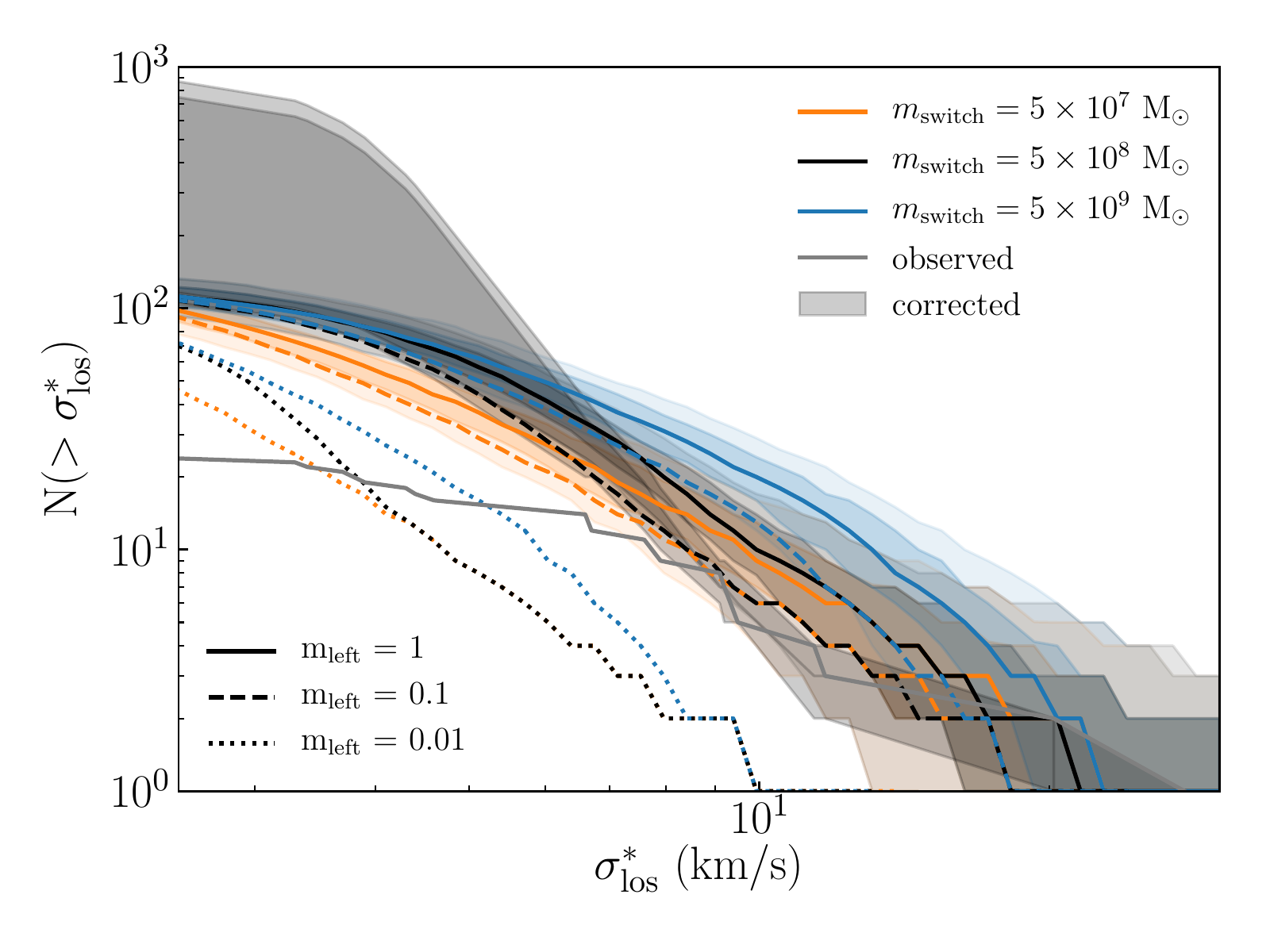}
    \caption{Same as Fig. \ref{fig:vfxns}, but modifying the switch mass $m_{\rm switch}$ at which cores vs cusps dominate.  The theoretical VF based on our fiducial switch mass of $m_{\rm switch} = 5 \times 10^8$ M$_\odot$ is shown in black, and variations of an order of magnitude higher/lower are shown in blue/orange.  Satellites with masses less than $m_{\rm switch}$ were assumed to have coreNFW density profiles, and NFW for more massive satellites.}
    \label{fig:vfxn-mswitch}
\end{figure}

\subsection{Switch mass between cores and cusps}
\label{appdx:theory_uncert:mswitch}

In \S\ref{sec:results:corecusp}, we showed that one method to alleviate the mismatch between the corrected and theory VFs was to introduce a switch between cored satellites at the high mass end and cuspy satellites at the low mass end, and that the best switch mass $m_{\rm switch} \sim 5 \times 10^8$ M$_\odot$.  In Fig. \ref{fig:vfxn-mswitch}, we show how the theoretical VF changes if $m_{\rm switch}$ changes by an order of magnitude.  As expected, a higher $m_{\rm switch}$ reproduces the large deviation at $\sim$10 km/s due to cusps in dwarfs at these scales, and a lower $m_{\rm switch}$ struggles to produce our most conservative completeness-corrections.  As such, we decided to adopt a fiducial $m_{\rm switch} = 5 \times 10^8$ M$_\odot$.

\begin{figure*}
    \centering
    \centerline{
    \includegraphics[width=0.5\textwidth]{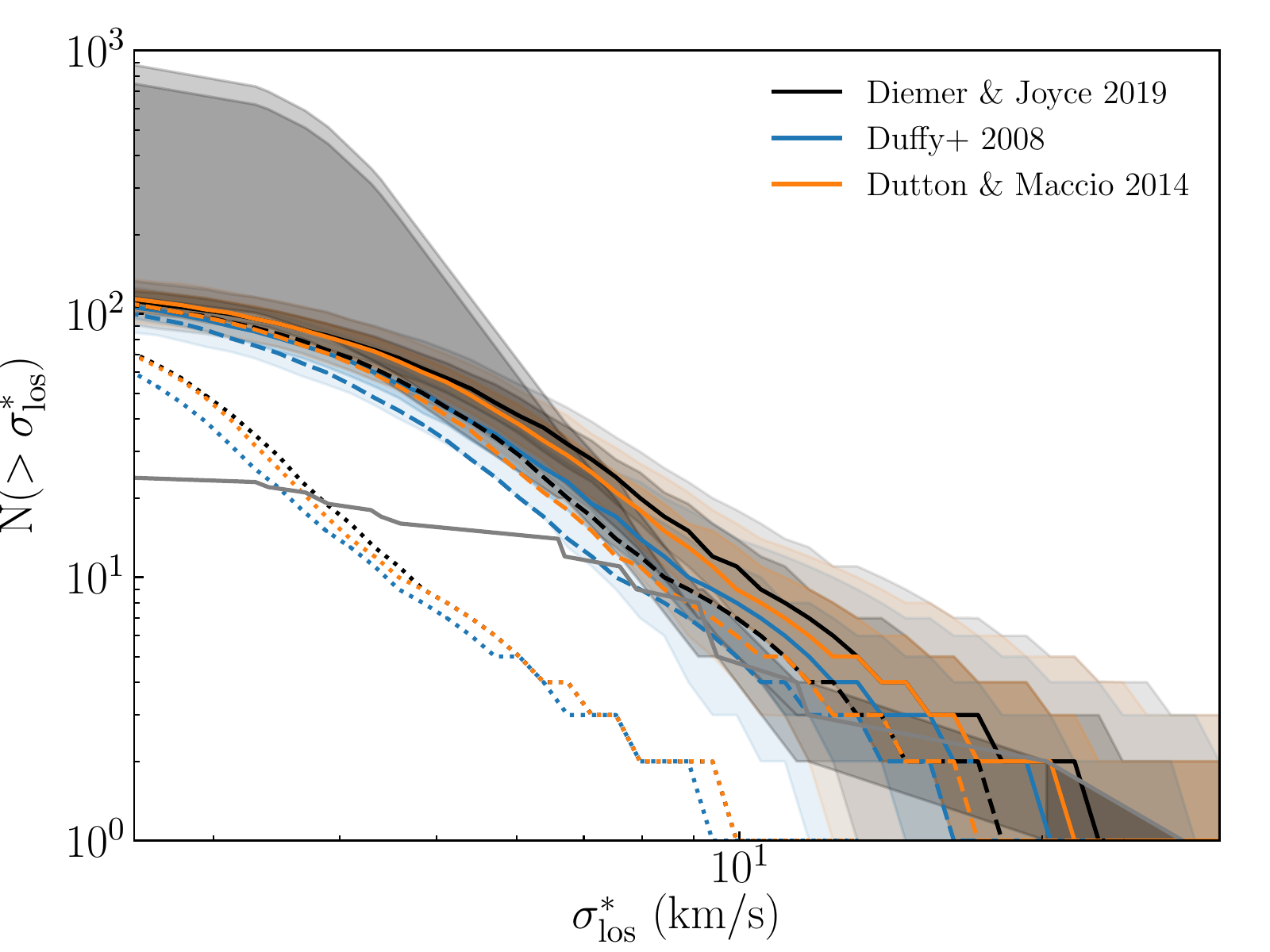}
    \includegraphics[width=0.5\textwidth]{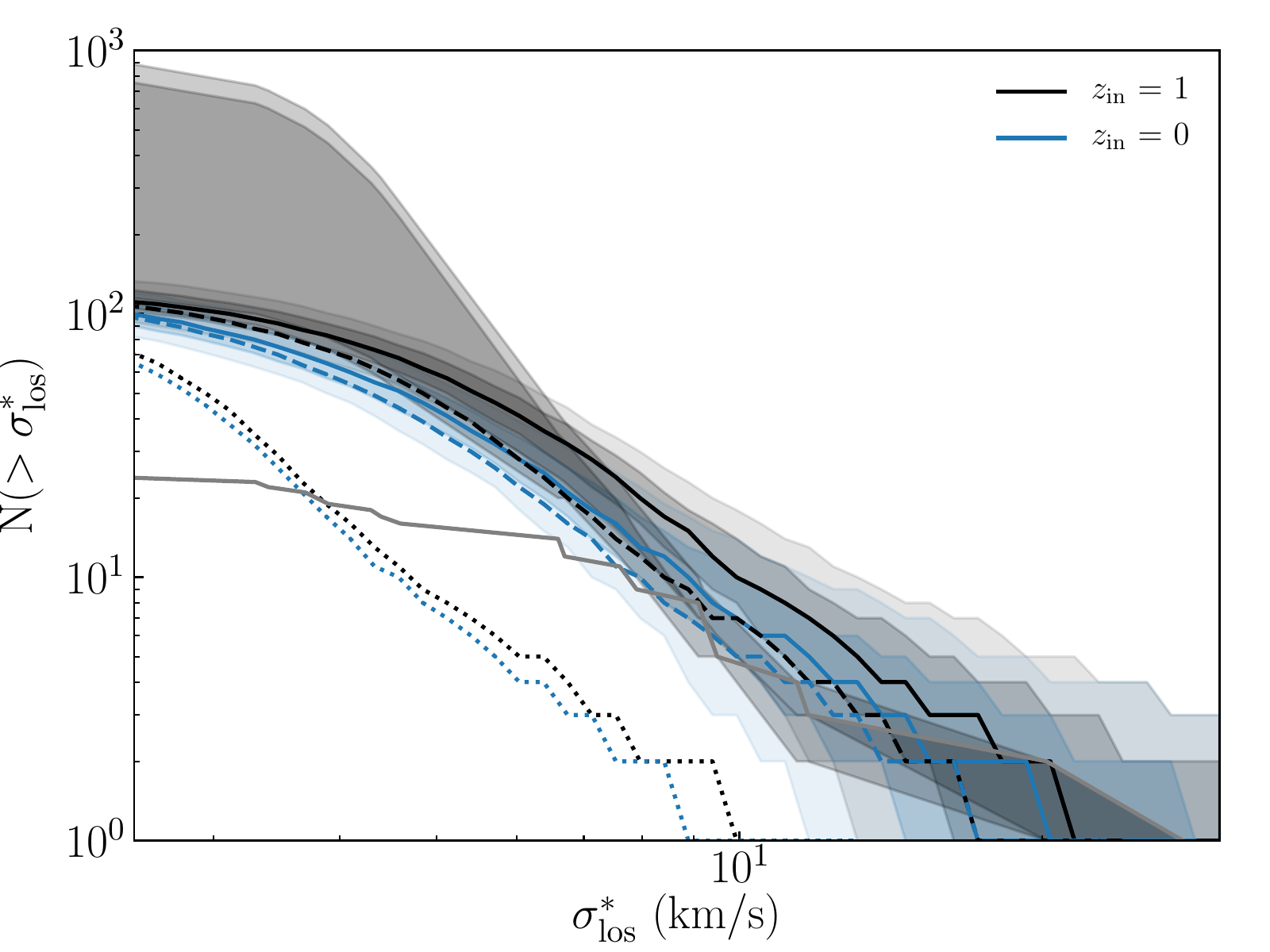}}
    \caption{Same as Fig. \ref{fig:vfxns}, but for theoretical uncertainties that do not significantly affect our conclusions:  (\emph{left}) mass-concentration relations and (\emph{right}) infall redshift $z_{\rm in}$.  In all panels, fiducial choices are plotted in black, and satellites with masses $M_{200} < 5 \times 10^8$ M$_\odot$ were assumed to have coreNFW density profiles, and NFW for more massive satellites.}
    \label{fig:vfxn_theory_uncert_addendum}
\end{figure*}

\subsection{Mass-concentration relation}
\label{sec:theoretical_uncert:mc_relation}

Significant scatter also exists in the mass-concentration relation.  In the left panel of Fig. \ref{fig:vfxn_theory_uncert_addendum}, we show our fiducial \citet{2019ApJ...871..168D} relation (black), which has a 0.16 dex 1$\sigma$ scatter \citep{2015ApJ...799..108D}. \citet{2019ApJ...871..168D} predict higher concentrations than other mass-concentration relations.  We additionally show \citet{2008MNRAS.390L..64D} (blue), which brackets the low end of concentrations.  We also show \citet{2014MNRAS.441.3359D}'s relation, which lies between the two.  Both of these lowers the theoretical VF relative to fiducial.  The plotted relations differ in assumed cosmologies; while \citet{2008MNRAS.390L..64D} was calibrated assuming a WMAP5 cosmology, we used the latest Planck 2018 cosmology with the \citet{2019ApJ...871..168D} relation, which is calibrated to work with any cosmology.  As shown in \citet{2019ApJ...871..168D}, the mass-concentration relation is a function of cosmological parameters (notably $\Omega_m$ and $\sigma_8$).

\subsection{Infall times}

The VF is somewhat sensitive to the infall time of the satellites due to the redshift dependence of the mass-concentration relation, which predicts higher concentrations at earlier redshifts. In Fig. \ref{fig:vfxn_theory_uncert_addendum}, we show how the VF changes if we assume an infall time of $z_{\rm in}$ = 0.  As before, have adopted a switch between cusps and cores at $5 \times 10^8$ M$_\odot$. The VF shifts downwards, underpredicting the most conservative completeness corrections at low velocities. 

While we have adopted a single infall time of $z_{\rm in} = 1$, the MW satellites have a range of infall times \citep{2012MNRAS.425..231R, 2017MNRAS.471.4894D, 2019arXiv190604180F}.  We predict that an exploration of plausible infall time distributions for the MW's satellite population will show that the assembly history of the MW will imprint itself in the VF of its satellites.

\begin{figure*}
    \centering
    \centerline{
    \includegraphics[width=0.5\textwidth]{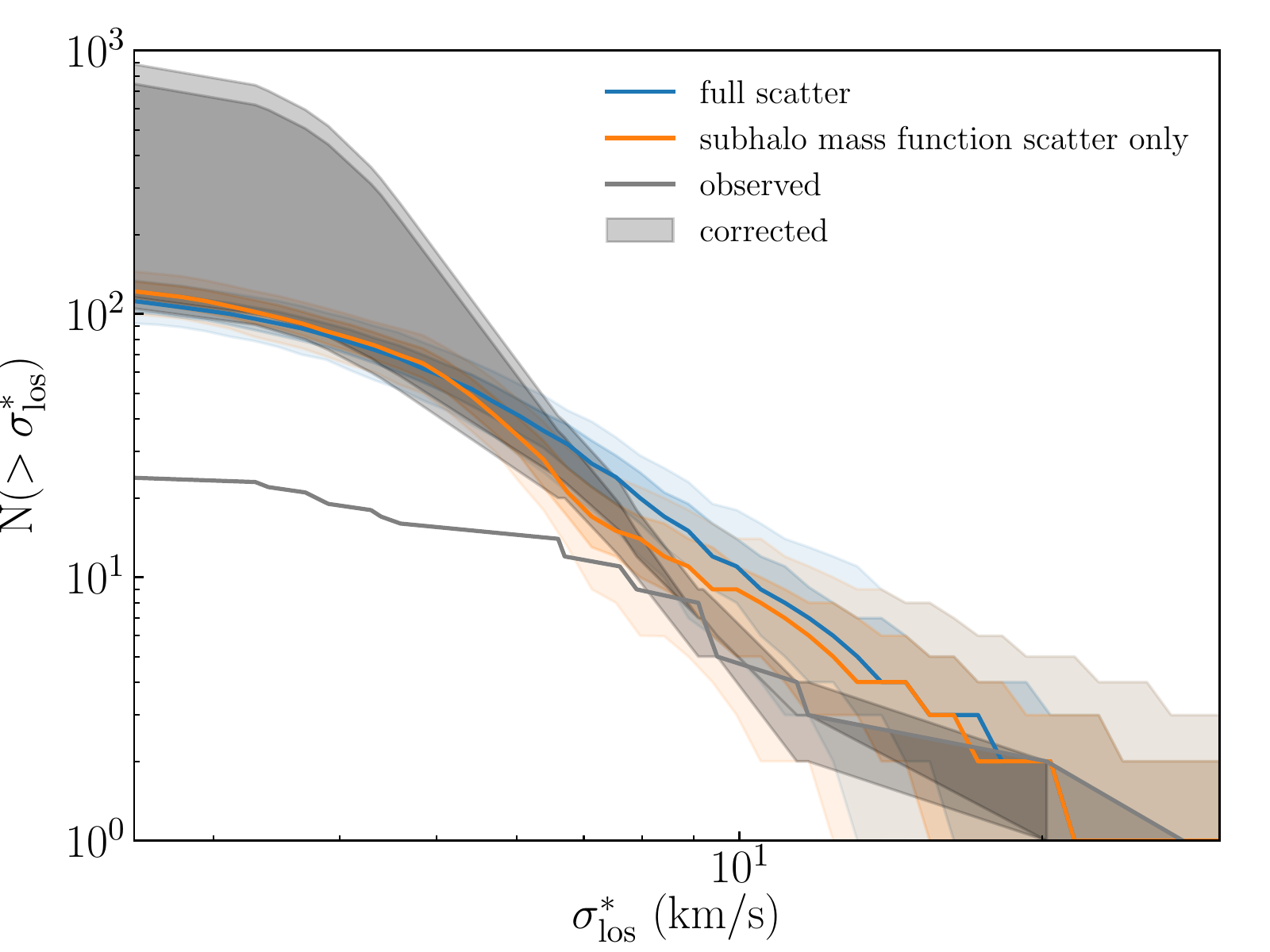}
    \includegraphics[width=0.5\textwidth]{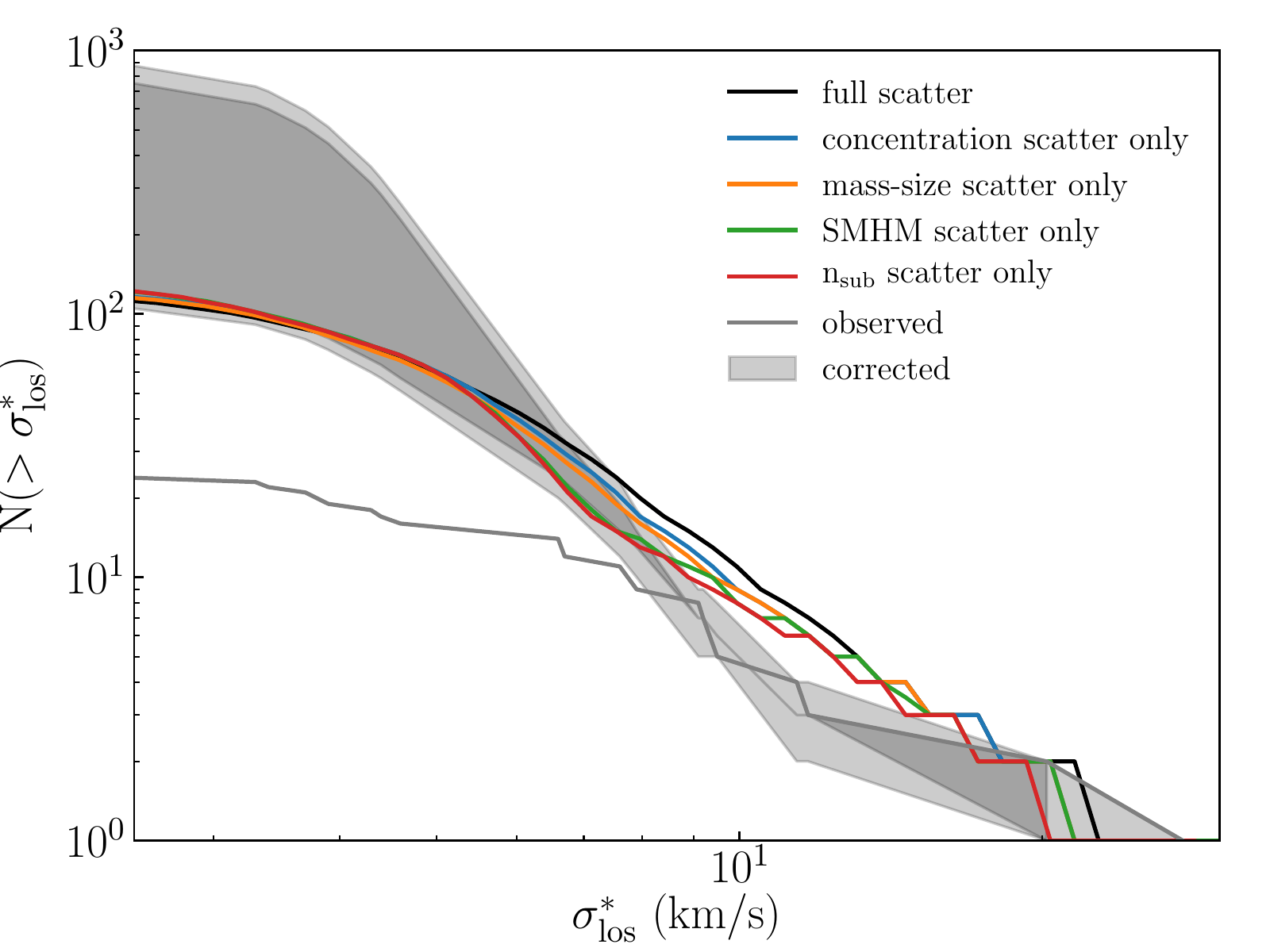}}
    \caption{The effect of scatter on the shape of the theoretical VF.  As usual, satellites with masses $M_{200} < 5 \times 10^8$ M$_\odot$ were assumed to have coreNFW density profiles, and NFW for more massive satellites.  (\emph{Left}) The theoretical VF including the fiducial full-scatter model in blue and only the scatter due to randomly selecting subhalos from the subhalo mass function (i.e. the halo-to-halo scatter) in orange.  The switch between the core-dominated and cusp-dominated regimes produces a sharper change in slope at intermediate $\sigma_{\rm los}^*$ in the theoretical VF if only the halo-to-halo scatter is considered.  (\emph{right}) The median theoretical VFs if one additional source of scatter in addition to the halo-to-halo scatter is included.  The sharper slope change disappears if the scatter in the mass-concentration or the stellar mass-size relations are included.}
    \label{fig:vfxn_theory_scatter}
\end{figure*}

\subsection{Relations with significant scatter and their impact on the theoretical VF}
\label{appdx:theory_uncert:scatter}

Lastly, we note that the size of the scatter can have significant effects on the shape of the theoretical VF.  In Fig. \ref{fig:vfxn_theory_scatter}, we show how the slope of the VF can change at intermediate $\sigma_{\rm los}^*$ if the levels of scatter in the scaling relations we have adopted are varied.  In the left panel, we show the fiducial, full-scatter model (blue) as well as the theoretical VF if the scatter in all relations except for that due to sampling the subhalo mass function are set to zero (orange).  A sharper change in slope due to the introduction of the scale-breaking switch between cusps and cores appears at intermediate $\sigma_{\rm los}^* \lesssim$ 10 km/s if the scatter is reduced.  

The smoothing of the sharp features introduced by the switch between cusps and cores is due to the large scatter in the mass-concentration and the stellar mass-size relations.  The right panel of Fig. \ref{fig:vfxn_theory_scatter} shows the median relations with the inclusion of one additional source of scatter in addition to the halo-to-halo scatter.  Including the scatter in the SMHM relation (green) or the number of subhalos of a MW-mass host (red) does not smooth over the change in slope.  In contrast, the mass-concentration (blue) and stellar mass-size (orange) relations do.  This highlights the sensitivity of the VF to the central densities of subhalos, which is dependent on the mass-concentration relation, as well as the half-light radii of satellites, which sets the radial extent out to which $\sigma_{\rm los}^*$ probes a satellite's density profile.

\bsp	
\label{lastpage}
\end{document}